\documentclass[conference]{IEEEtran}
\IEEEoverridecommandlockouts
\usepackage[nospace,sort]{cite}
\usepackage{amsmath,amssymb,amsfonts}
\usepackage{amsthm}
\usepackage{graphicx}
\usepackage{textcomp}
\usepackage{xcolor}
\def\BibTeX{{\rm B\kern-.05em{\sc i\kern-.025em b}\kern-.08em
    T\kern-.1667em\lower.7ex\hbox{E}\kern-.125emX}}

\usepackage[ruled,vlined,linesnumbered]{algorithm2e}
\usepackage{booktabs}
\usepackage{url}
\usepackage[flushleft]{threeparttable}
\usepackage{hyperref}
\usepackage{amsbsy}

\ifCLASSOPTIONcompsoc
\usepackage[caption=false, font=normalsize, labelfont=sf, textfont=sf]{subfig}
\else
\usepackage[caption=false, font=footnotesize]{subfig}
\fi

\begin{document}

\title{Dynamic Similarity Search on Integer Sketches}

\author{\IEEEauthorblockN{Shunsuke Kanda}
\IEEEauthorblockA{\textit{RIKEN Center for Advanced Intelligence Project}\\
Tokyo, Japan \\
shunsuke.kanda@riken.jp}
\and
\IEEEauthorblockN{Yasuo Tabei}
\IEEEauthorblockA{\textit{RIKEN Center for Advanced Intelligence Project}\\
Tokyo, Japan \\
yasuo.tabei@riken.jp}}

\newcommand{\Times}[1]{#1$\times$}

\newcommand{\FuncFont}[1]{\textsf{#1}}

\newcommand{\KeyS}{\FuncFont{Key}}
\newcommand{\PtrS}{\FuncFont{Ptr}}
\newcommand{\IdxD}{\FuncFont{Idx}}
\newcommand{\PtrD}{\FuncFont{Ptr}}
\newcommand{\PtrF}{\FuncFont{Ptr}}

\newcommand{\LUT}{\FuncFont{A}}
\newcommand{\HDT}{\FuncFont{H}}

\newcommand{\NIL}{\FuncFont{nullptr}}

\newcommand{\LS}{\FuncFont{LinearScan}}
\newcommand{\BF}{\FuncFont{BruteForce}}

\newcommand{\NS}{\FuncFont{NodeS}}
\newcommand{\ND}{\FuncFont{NodeD}}
\newcommand{\NF}{\FuncFont{NodeF}}

\newcommand{\InsertFn}{\FuncFont{Insert}}
\newcommand{\SearchFn}{\FuncFont{Search}}

\newcommand{\ASFTP}{DyFT$^+$}

\newcommand{\Ceil}[1]{\lceil{#1}\rceil}
\newcommand{\Floor}[1]{\lfloor{#1}\rfloor}

\newcommand{\qref}[1]{Eq. (\ref{#1})}
\newcommand{\fref}[1]{Figure \ref{#1}}
\newcommand{\ffref}[2]{Figures \ref{#1} and \ref{#2}}
\newcommand{\ftfref}[2]{Figures \ref{#1}--\ref{#2}}
\newcommand{\tref}[1]{Table \ref{#1}}
\newcommand{\ttref}[2]{Tables \ref{#1} and \ref{#2}}
\newcommand{\sref}[1]{Section \ref{#1}}
\newcommand{\ssref}[2]{Sections \ref{#1} and \ref{#2}}
\newcommand{\gref}[1]{Algorithm \ref{#1}}
\newcommand{\pref}[1]{Property \ref{#1}}
\newcommand{\aref}[1]{Appendix \ref{#1}}

\newcommand\mycommfont[1]{\footnotesize\rmfamily\textcolor[rgb]{0,0.46,0.73}{#1}}
\SetCommentSty{mycommfont}
\SetArgSty{\rmdefault}
\SetFuncSty{\rmdefault}
\SetKwInOut{Input}{Input}
\SetKwInOut{Output}{Output}
\SetKwProg{Fn}{function}{}{}
\SetKwProg{Pc}{procedure}{}{}
\SetKwComment{tcp}{{$\triangleright$} }{}

\newtheorem{problem}{Problem}

\maketitle

\begin{abstract}
Similarity-preserving hashing is a core technique for fast similarity searches, and it randomly maps data points in a metric space to strings of discrete symbols (i.e., sketches) in the Hamming space. While traditional hashing techniques produce binary sketches, recent ones produce integer sketches for preserving various similarity measures. However, most similarity search methods are designed for binary sketches and inefficient for integer sketches. Moreover, most methods are either inapplicable or inefficient for dynamic datasets, although modern real-world datasets are updated over time. We propose dynamic filter trie (DyFT), a dynamic similarity search method for both binary and integer sketches. An extensive experimental analysis using large real-world datasets shows that DyFT performs superiorly with respect to scalability, time performance, and memory efficiency. For example, on a huge dataset of 216 million data points, DyFT performs a similarity search 6,000 times faster than a state-of-the-art method while reducing to one-thirteenth in memory.
\end{abstract}

\section{Introduction}

Similarity search of vectorial data in databases has been a fundamental task in recent data analysis and has various applications such as near duplicate detection in a collection of web pages~\cite{henzinger2006finding}, context-based retrieval in images~\cite{song2013inter}, and functional analysis of molecules~\cite{ito2012possum}.
In recent years, databases for these applications have become larger and the vectorial data of these databases have also become dimensionally higher, making it difficult to apply existing similarity search methods to such large databases.
Therefore, it is necessary to develop much more powerful similarity search methods to analyze large databases efficiently.

Similarity-preserving hashing is a powerful technique that approximates a similarity measure by randomly mapping data points in a metric space to strings of discrete symbols (i.e., \emph{sketches}) in the Hamming space.
Similarity search problems for various similarity measures can be approximately solved as the \emph{Hamming distance problem} for sketches 
(i.e., computations of the number of positions at which the corresponding integers between two sketches are different).
Thus far, many hashing techniques producing binary sketches have been developed as reviewed in \cite{cao2017binary};
accordingly, quite a few similarity search methods especially for binary sketches have been proposed for decades (e.g., \cite{gog2016fast,qin2019generalizing,norouzi2014fast,eghbali2019online}).
In recent years, many types of hashing techniques intending to produce integer sketches have been developed for various similarity measures.
Examples are $b$-bit minwise hashing for Jaccard similarity \cite{li2010b}, 0-bit consistent weighted sampling (CWS) for min-max kernels \cite{li20150}, and 0-bit CWS for generalized min-max kernels \cite{li2017linearized}.
There is a strong need to develop efficient solutions for the \emph{general Hamming distance problem} for not only binary sketches but also integer sketches;
however, few similarity search methods designed on the general problem have been proposed \cite{zhang2013hmsearch,kanda2019b,kanda2020tstat}.

Modern real-world datasets are updated over time, which we shall call dynamic setting. 
For example, search engines often have a large number of new web pages containing images and text data, which arrive in the data center every day.
Dynamic similarity search methods that can efficiently perform insertions of new data points to a dataset and deletions of data points from the dataset are essential in modern data mining and information retrieval.
However, most of state-of-the-art methods have drawbacks: (i) limitations to static settings \cite{gog2016fast,qin2019generalizing,kanda2019b,kanda2020tstat} or (ii) inefficiency in dynamic settings \cite{norouzi2014fast,zhang2013hmsearch}.
Although Eghbali et al. \cite{eghbali2019online} recently proposed Hamming weight tree (HWT) to address this problem, it is applicable only to binary sketches and its performance is degraded for large datasets.
Consequently, an important open challenge is to develop a fast, scalable, and dynamic similarity search method for the general Hamming distance problem. 

Our main contributions in this paper are as follows:

\begin{itemize}
\item We propose \emph{dynamic filter trie (DyFT)}, a dynamic similarity search method for both binary and integer sketches using an edge-labeled tree called \emph{trie} \cite{fredkin1960trie}. DyFT grows the data structure based on a search cost model to maintain fast similarity searches. It also reduces on memory consumption by omitting redundant trie nodes (\sref{sect:asft}).
\item We design an implementation for DyFT, called \emph{modified adaptive radix tree (MART)}, in which the data structure changes adaptively depending on the configuration of DyFT nodes. MART always enables DyFT to perform well for any input parameter of similarity-preserving hashing (\sref{sect:impl}).
\item We present an extensive experimental analysis that shows DyFT performs superiorly compared to state-of-the-art similarity search methods for both binary and integer sketches with respect to scalability, time performance, and memory efficiency. For example, on a huge dataset of 216 million sketches, DyFT performs a similarity search 6,000 times faster than HWT, while reducing to one-thirteenth in memory (\sref{sect:ex}).
\end{itemize}

\begin{table*}[tb]
\centering
\begin{threeparttable}
\caption{
Summary of Similarity Search Methods.
}
\label{tab:review}
\begin{tablenotes}
\begin{tabular}{rlllll} 
\toprule
\textbf{Method} & \textbf{Sketch type} & \textbf{Data structure} & \textbf{Search time} & \textbf{Update time} & \textbf{Memory} \\
\midrule
MIH \cite{norouzi2014fast} & Binary & Hash table & $O(q (m/q)^{{r/q}} \cdot \max(1, n / 2^{m/q} )) + V_\mathrm{mih}$ & $O(q)$ & $O(mn)$ \\
HWT \cite{eghbali2019online} & Binary & Search tree & $O(m \log m (\log n)^{4r}) +  V_\mathrm{hwt}$ & $O(m \log m)$ & $O(mn)$ \\
HmSearch (HSV) \cite{zhang2013hmsearch} & Integer & Hash table & $O(r \cdot \max(1,mn\sigma /\sigma^{m-{2m}/{r}})) + V_\mathrm{hsv}$ & $O(m^2 \sigma /r)$ & $O(mn\sigma )$ \\
HmSearch (HSD) \cite{zhang2013hmsearch} & Integer & Hash table & $O(m \cdot \max(1,mn/\sigma^{m-{2m}/{r}+1})) + V_\mathrm{hsd}$ & $O(m^2/r)$ & $O(mn)$ \\
GV \cite{gog2016fast} & Integer & Hash table & $O((m+r) \cdot \max(1,n /\sigma^{2m/(r+2)})) + V_\mathrm{gv}$ & $O(m)$ & $O(mn)$ \\
\midrule[.03em]
DyFT (this study) & Integer & MART & $O(m^{r+2}) +  V_\mathrm{dyft}$ & $O(m)$ & $O(mn)$ \\
DyFT$^{+}$ (this study) & Integer & MART & $O(q (m/q)^{{r/q}+2}) +  V_\mathrm{dyft+}$ & $O(m)$ & $O(mn)$ \\
\bottomrule
\end{tabular}
\footnotesize
\item \emph{Note:} $V$ is verification time for candidates obtained from each similarity search method.
\end{tablenotes}
\end{threeparttable}
\end{table*}

\section{Problem Statement}
Sketch $x$ of length $m$ is an $m$-dimensional vector of non-negative integers from alphabet $\Sigma = \{0,1, \dots, \sigma - 1 \}$ of alphabet size $\sigma$, i.e., $x \in \Sigma^m$.
The $i$-th element of $x$ is denoted by $x[i]$.
The Hamming distance between sketches $x$ and $y$ is the number of positions at which the corresponding elements are different, formally defined as
\begin{equation*}
H(x,y) = \sum^{m}_{i=1} \left\{
\begin{array}{cl}
1 & (x[i] \neq y[i]) \\
0 & (x[i] = y[i])
\end{array}
\right.
.
\end{equation*}
We assume $m = O(w)$ for word size $w$.
Then, $H(x,y)$ can be computed in $O(\log\sigma)$ time by performing $\Ceil{\log_2 \sigma}$ sets of bitwise XOR and popcount operations \cite{zhang2013hmsearch}.

A database $X=\{x_1,x_2, \dots ,x_n\}$ is a dynamic set consisting of $n$ sketches, and it supports the insertion of a new sketch $x_i$ and deletion of sketch $x_i$.
The general Hamming distance problem for a given sketch $y$ and radius $r$ is to find all the sketches whose Hamming distance to sketch $y$ in $X$ is at most $r$, i.e., $R = \{x_i \in X : H(x_i,y) \leq r \}$.

\section{Literature Review}
\label{sect:review}

Many similarity search methods on Hamming distance have been proposed for decades.
Several recent studies have focused on static settings \cite{gog2016fast,qin2019generalizing,kanda2019b,kanda2020tstat}.
Theoretical aspects have also been argued \cite{belazzougui2012compressed,cole2004dictionary,chan2010compressed}.
In this section, we briefly review state-of-the-art similarity search methods for binary and integer sketches, and they are applicable to dynamic datasets. 
\tref{tab:review} summarizes state-of-the-arts.

A seminal work for binary sketches is {multi-index hashing (MIH)} developed by Norouzi et al. \cite{norouzi2014fast}.
MIH is based on the \emph{multi-index} approach \cite{greene1994multi} and it enables quick similarity searches even with large $r$.
A key observation is that two similar sketches must have similar parts.
Thus, MIH partitions each sketch into $q$ blocks of short sketches and builds $q$ hash tables from the short sketches in each block.
The similarity search first obtains a set of candidate solutions $R' \supseteq R$ by retrieving each hash table with small radius $\Floor{r/q}$ and then removes false positives from $R'$ by computing the Hamming distances.

The number of blocks offering the best search performance is determined by the configuration of the dataset.
Norouzi et al. \cite{norouzi2014fast} empirically demonstrated that the best performance of MIH is often achieved when $q = m/\log_2 n$.
They also showed that setting $q$ to a number apart from $m/\log_2 n$ significantly degrades performance.
Thus, MIH is unsuitable for dynamic problem settings where database size $n$ varies.

{Hamming weight tree (HWT)} developed by Eghbali et al. \cite{eghbali2019online} is a state-of-the-art similarity search method to solve the issue of MIH.
Instead of using hash tables, HWT uses a search tree constructed based on Hamming weight (i.e., the number of ones appearing in a binary sketch).
However, the similarity search takes $O(m \log m (\log n)^{4r})$ time and slows down dramatically for a large database of $n$.
In addition, those similarity search methods were designed for binary sketches, and they are not necessarily suitable for integer sketches. 

HmSearch developed by Zhang et al. \cite{zhang2013hmsearch} is a multi-index similarity search method designed for integer sketches.
HmSearch reduces the general Hamming distance problem with radius $r$ to small problems with radius one by tuning the number of blocks and preregistering candidate solutions in hash tables.
However, this approach preregistering candidate solutions consumes a large amount of memory and requires a large amount of update time.

Gog and Venturini \cite{gog2016fast} proposed an idea that defines $\Floor{r/2} + 1$ blocks to produce small problems with radius one and bypasses preregistering candidate solutions stored in hash tables. 
They presented a simple variant of HmSearch, which is referred to as GV in this paper.
The similarity search is performed with the same algorithm as that of MIH.
Thus, GV can be considered as a simple modification of MIH for integer sketches and has the same issue as MIH, which causes inefficiency in dynamic problem settings where $n$ is variable.
Moreover, GV's search speed was much slower than HmSearch's, as experimentally demonstrated in \sref{sect:ex}.

Despite the importance of dynamic similarity search methods for the general Hamming distance problem, there is no efficient method.
The main reason is that most methods rely on the multi-index approach using hash tables, which require setting the appropriate number of blocks depending on variable parameter $n$.
Although HWT attempts to address that issue using a tree structure, it is inefficient for large datasets and is inapplicable to integer sketches.

\section{Dynamic Filter Trie}
\label{sect:asft}

DyFT is a dynamic similarity search method for the general Hamming distance problem.
As with HWT, DyFT is built on a tree-based data structure.
In contrast to HWT, DyFT employs a \emph{trie} data structure \cite{fredkin1960trie}, which enables quick similarity searches for integer sketches.
In this section, we first introduce the trie data structure and the design motivation of DyFT;
Then, we present DyFT's data structure and complexity analyses.

\subsection{Preliminaries}
\label{sect:asft:pre}

Trie is an edge-labeled tree storing a set of sketches.
Each node is associated with the common prefix of a subset of the sketches, and each leaf is associated with a particular sketch in the database.
Each edge has an integer organizing sketches as a label.
All outgoing edges of an inner node are labeled with distinct integers.
The downgoing path from the root to each leaf corresponds to the sketch associated to the leaf.

The exact search for a given sketch $y$ traverses trie nodes from the root by using the integers of $y$.
If we reach a leaf, $y$ is stored in the trie.
A simple extension of the exact search implements the similarity search with radius $r$.
The similarity search traverses trie nodes from the root by using the integers of $y$ with at most $r$ errors allowed.
In other words, we count the number of errors from the root to each node $v$ visited in the traversal and, if the number exceeds $r$, stop traversing down to all the descendants under $v$.
The solution $R$ is the set of all sketches associated with leaves reachable within $r$ errors.
A more specific description of the similarity search algorithm using trie is presented in \cite[Sect. IV-B]{kanda2019b}.
The similarity search can prune unnecessary portions of the search space and can be quickly performed for a small radius $r$.
The time complexity is $O(m^{r+2})$ \cite{arslan2002dictionary}.\footnote{Although Arslan and Eğecioğlu \cite{arslan2002dictionary} derived the complexity assuming $\sigma = 2$, it does not vary for any $\sigma$.}

Each inner node in a trie is implemented as a mapping from edge labels to child pointers.
A trie storing a large database $X$ maintains many pointers and consumes a large amount of memory.
A well-known technique for substantially reducing memory consumption is to omit nodes around leaves.
Thus far, a number of memory-efficient trie data structures have been developed by leveraging this technique, e.g., \cite{askitis2010engineering,heinz2002burst,zhang2018surf,leis2013adaptive}.
However, these data structures were designed for exact string searches and inefficient for similarity searches.

There is no dynamic and scalable trie data structure for similarity searches.
In the remainder of this section, we present DyFT, which omits many nodes while maintaining fast similarity searches.
DyFT's performance also depends on the implementation of the mapping for each inner node.
In \sref{sect:impl}, we introduce an efficient node implementation for DyFT.

\subsection{Approach}

\begin{figure}[tb]
\centering
\includegraphics[scale=0.5]{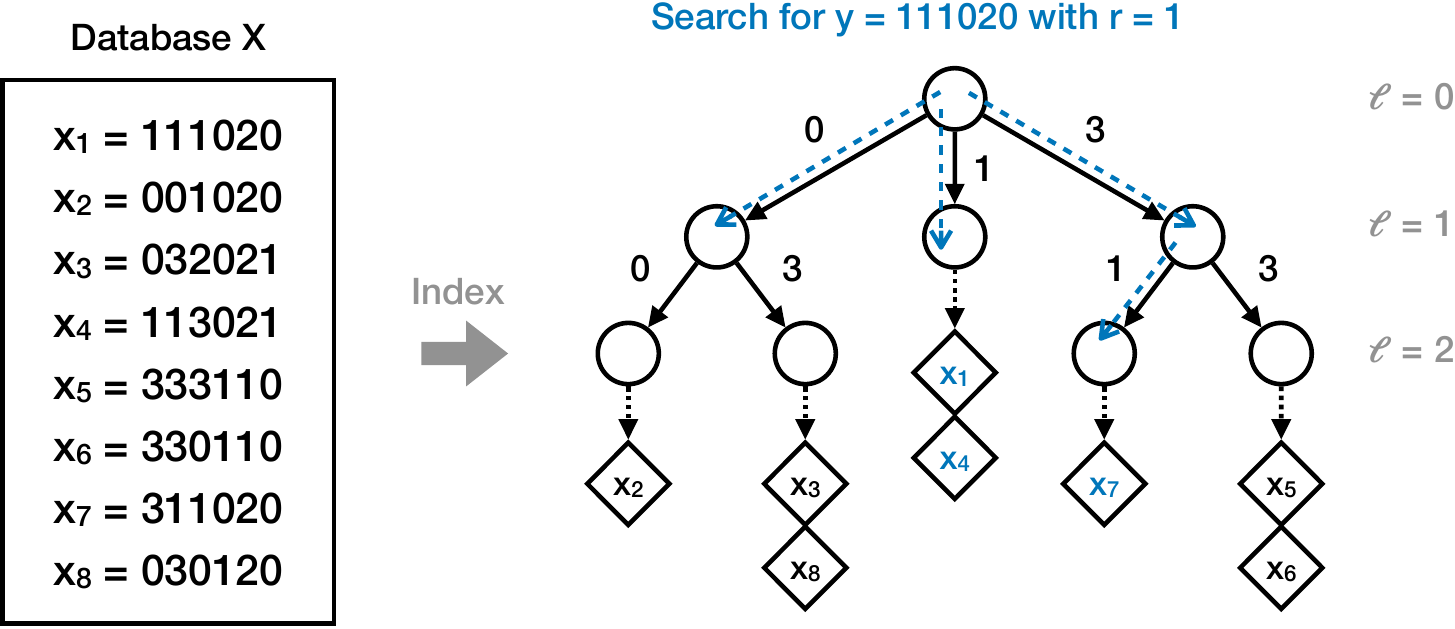}
\caption{
Example of DyFT for database $X$.
Search for $y$ with $r = 1$ traverses nodes along blue dashed arrows and reaches the posting lists containing $x_1$, $x_4$, and $x_7$.
The solution $R = \{ x_1, x_7 \}$ is obtained by verifying $H(x_1,y) = 0$, $H(x_4,y) = 2$, and $H(x_7,y) = 1$.
}
\label{fig:asft}
\end{figure}

The basic idea is to allow false positives and store only some of trie nodes around the root.
In other words, DyFT exploits the trie search algorithm for \emph{filtering out} dissimilar sketches and aims to obtain solution candidates.
\fref{fig:asft} shows an example of DyFT for eight sketches.
A leaf $v$ at level $\ell$ reached by sub-sketch $x' \in \Sigma^{\ell}$ is associated with all sketches in $X$ starting with $x'$.
For example, in \fref{fig:asft}, the leaf reached by ``03'' is associated with sketches $x_3$ and $x_8$ starting with ``03''.
Every leaf $v$ has the posting list of associated sketches.
We denote the posting list by $L_v$ and its length by $|L_v|$.

The similarity search for given $y$ and $r$ traverses DyFT nodes in the aforementioned manner.
For a leaf $v$ reached within $r$ errors, each sketch $x_i \in L_v$ is verified by checking whether $H(x_i,y) \leq r$.
\fref{fig:asft} shows a search example, and \gref{algo:asft} shows the search algorithm.

\begin{algorithm}[tb]
\footnotesize

\DontPrintSemicolon

\SetKwFunction{KwFnInsert}{\FuncFont{Insert}}
\SetKwFunction{KwFnSearch}{\FuncFont{Search}}

\caption{Search and insertion algorithms of DyFT}
\label{algo:asft}

\Fn{\KwFnSearch{$y,r$}}{
\tcp{Traverse DyFT nodes using a stack}
$R \gets \emptyset, S_{\mathrm{stack}} \gets \{ (v_{\mathrm{root}}, 0, r) \}$\tcp*[r]{$v_{\mathrm{root}}$: DyFT's root}
\While(){$S_{\mathrm{stack}} \neq \emptyset$}{
	\tcp{$r'$ is $r$ minus the number of errors at node $v$ of level $\ell$}
	Pop back $(v, \ell, r')$ from $S_{\mathrm{stack}}$\;
	\If(\tcp*[f]{Verify the candidates in $L_v$}){$v$ is a leaf}{
		\For{$x_i \in L_v$}{
			\lIf{$H(x_i,y) \leq r$}{
				Append $x_i$ to $R$
			}
		}
		\textbf{continue}\;
	}
	\uIf(\tcp*[f]{Check all the children of $v$}){$r' > 0$}{\label{line:asft:case1_beg}
		\For{$u$ in the set of children of $v$}{\label{line:asft:scanfor_beg}
			\uIf{$u$'s edge label is $y[\ell+1]$}{
				Push back $(u, \ell+1, r')$ to $S_{\mathrm{stack}}$\;
			}
			\Else{
				Push back $(u, \ell+1, r' - 1)$ to $S_{\mathrm{stack}}$\;
			}
		}\label{line:asft:scanfor_end}
	}\label{line:asft:case1_end}
	\ElseIf(\tcp*[f]{Look up the child of $v$}){$r' = 0$}{\label{line:asft:case2_beg}
		\If{$v$ has a child $u$ with edge $y[\ell+1]$}{\label{line:asft:findif_beg}
			Push back $(u, \ell+1, r')$ to $S_{\mathrm{stack}}$\;
		}\label{line:asft:findif_end}
	}\label{line:asft:case2_end}
}
\KwRet $R$\;
}

\BlankLine

\Pc{\KwFnInsert{$y$}}{
$v \gets v_{\mathrm{root}}, \ell \gets 1$\;%
\While(\tcp*[f]{Traverse DyFT nodes}){$v$ is not a leaf}{
    \If(){$v$ does not have a child with edge $y[\ell]$}{
        Insert a new child to $v$ with edge $y[\ell]$\;
    }
    $v \gets$ the child of $v$ with edge $y[\ell], \ell \gets \ell + 1$\;
}
Append $y$ to $L_v$\;
\If(\tcp*[f]{Split $v$ and create new leaves from $v$}){$|L_v| > \tau$}{
    \For{$c \in \{ x_i[\ell] \mid x_i \in L_v \}$}{
        Insert new leaf $u$ from $v$ with edge label $c$\;
        Create new posting list $L_{u} = \{ x_i \in L_v \mid x_i[\ell] = c \}$\;
    }
	Remove the old posting list $L_v$\;
}
}
\end{algorithm}

\begin{figure}[tb]
\centering
\includegraphics[scale=0.5]{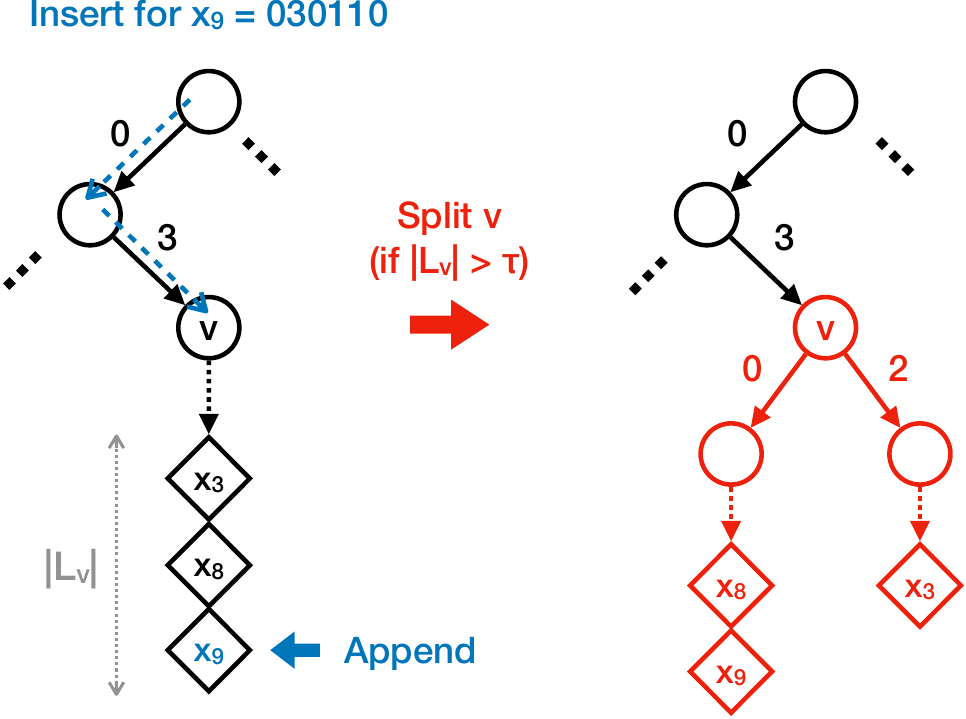}
\caption{
Example of inserting $x_9$ to $L_v$ in \fref{fig:asft}.
If $|L_v|$ is more than $\tau$, we split $v$ into two leaves.
}
\label{fig:split}
\end{figure}

We now present the insertion algorithm.
Initially, the DyFT structure for an empty $X$ consists only of the root with an empty posting list.
Given a sketch $x_i$, we traverse DyFT nodes using $x_i$ and visit the deepest reachable node $v$.
If $v$ is an inner one, we insert a new leaf $u$ from $v$ and associate a new posting list $L_u$ storing $x_i$.
If $v$ is a leaf, we append $x_i$ to $L_v$;
Then, DyFT determines whether leaf $v$ should be \emph{split}.
If $|L_v|$ is longer than a threshold $\tau$, we create new leaves from $v$ and split $L_v$ into disjoint short lists (see \fref{fig:split}).
\gref{algo:asft} shows the insertion algorithm.

The deletion algorithm is symmetrical to that of insertion.
We remove $x_i$ from $L_v$ for leaf $v$ reached by $x_i$.
If $L_v$ becomes empty, we remove leaf $v$ from DyFT.

\subsection{Optimal Threshold}
\label{sect:asft:opt}

The search performance of DyFT is affected by threshold $\tau$.
If $\tau$ is large, the verification time for $L_v$ becomes large.
If $\tau$ is small, DyFT defines many nodes and the traversal time becomes large.
Thus, we need to set a reasonable value of $\tau$.
Such a reasonable value of $\tau$ can be determined according to the configuration of $X$ and given parameters such as $n$, $\sigma$, and $r$; however, it is impossible to search such a reasonable value in dynamic settings.
To address this issue, we first construct a search cost model assuming that sketches are uniformly distributed in the Hamming space and then determine an optimal threshold $\tau^{*}$ minimizing the search cost.

By fixing $r$ and $\sigma$, we consider the \emph{reach probability} for node $v$ at level $\ell$, which is the probability to reach $v$ within $r$ errors using a random sketch $x \in \Sigma^{\ell}$ from a uniform distribution.
Let $v$ be traversed from the root node using sketch $\phi(v) \in \Sigma^{\ell}$.
The set of all sketches reachable to $v$ within $r$ errors is $\{x \in \Sigma^{\ell} : H(x, \phi(v)) \leq r \}$ whose cardinality is
\begin{equation*}
N(\ell) = \sum_{k=0}^{r} \binom{\ell}{k} ( \sigma - 1 )^k .
\end{equation*}
As the number of all possible sketches of length $\ell$ is $\sigma^{\ell}$, the reach probability of a node at level $\ell$ is 
\begin{equation*}
P(\ell) = \left\{
\begin{array}{cl}
1 & (\ell \leq r ) \\
N(\ell) / \sigma^{\ell} & (\ell > r)
\end{array}
\right.
.
\end{equation*}
It holds that $P(\ell) > P(\ell+1)$ for $\ell \geq r$.

We define the search cost of node $v$ at level $\ell$ for random sketch $x \in \Sigma^{\ell}$ by multiplying the reach probability by the computational cost.
When we visit an inner node $v$ at level $\ell$ during similarity search, we try to descend to the children of $v$.
Then, we have two cases whether (i) $H(x,\phi(v)) < r$ or (ii) $H(x,\phi(v)) = r$.
In case (i), we check all the children in $O(\sigma)$ time (Lines \ref{line:asft:scanfor_beg}--\ref{line:asft:scanfor_end} in \gref{algo:asft}).
In case (ii), we look up the child in $O(1)$ time (Lines \ref{line:asft:findif_beg}--\ref{line:asft:findif_end} in \gref{algo:asft}).
Case (ii) occurs for sketches in $\{x \in \Sigma^{\ell} : H(x, \phi(v)) = r \}$ whose cardinality is
\begin{equation*}
N_{2}(\ell) = \binom{\ell}{r} ( \sigma - 1 )^r .
\end{equation*}
The occurrence probability of case (ii) is ${N_{2}(\ell)} / {N(\ell)}$, and the computational cost of $v$ is 
\begin{equation*}
F_{\mathrm{in}}(\ell) = \left( 1 - \frac{N_{2}(\ell)}{N(\ell)} \right) \times \sigma + \frac{N_{2}(\ell)}{N(\ell)}.
\end{equation*}
Thus, the search cost of inner node $v$ at level $\ell$ is 
$C_{\mathrm{in}}(v) = P(\ell) \times F_{\mathrm{in}}(\ell)$.
When we visit a leaf $v$ at level $\ell$, we verify all sketches associated with $L_v$, and the search cost is
$C_{\mathrm{leaf}}(v) = P(\ell) \times |L_v| \times \Ceil{\log_2 \sigma}$.

\begin{figure}[tb]
\centering
\includegraphics[width=40mm]{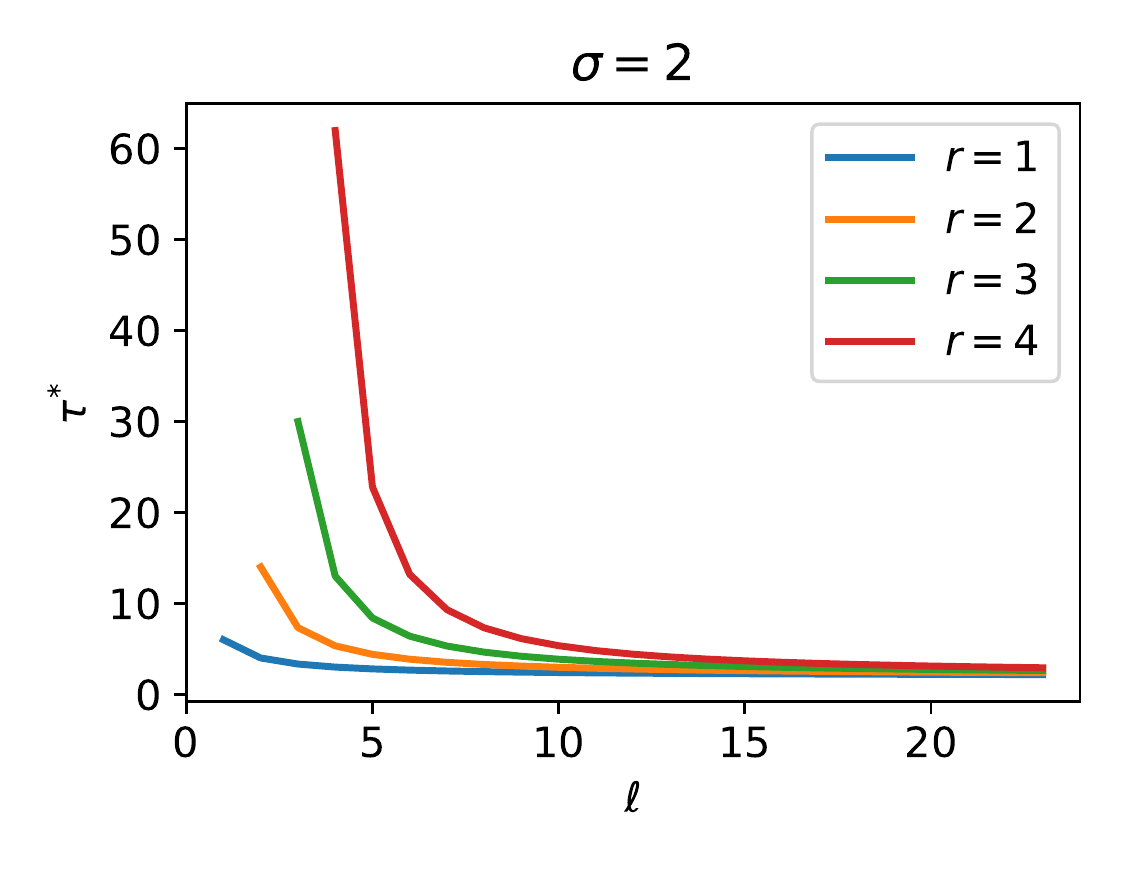}
\includegraphics[width=40mm]{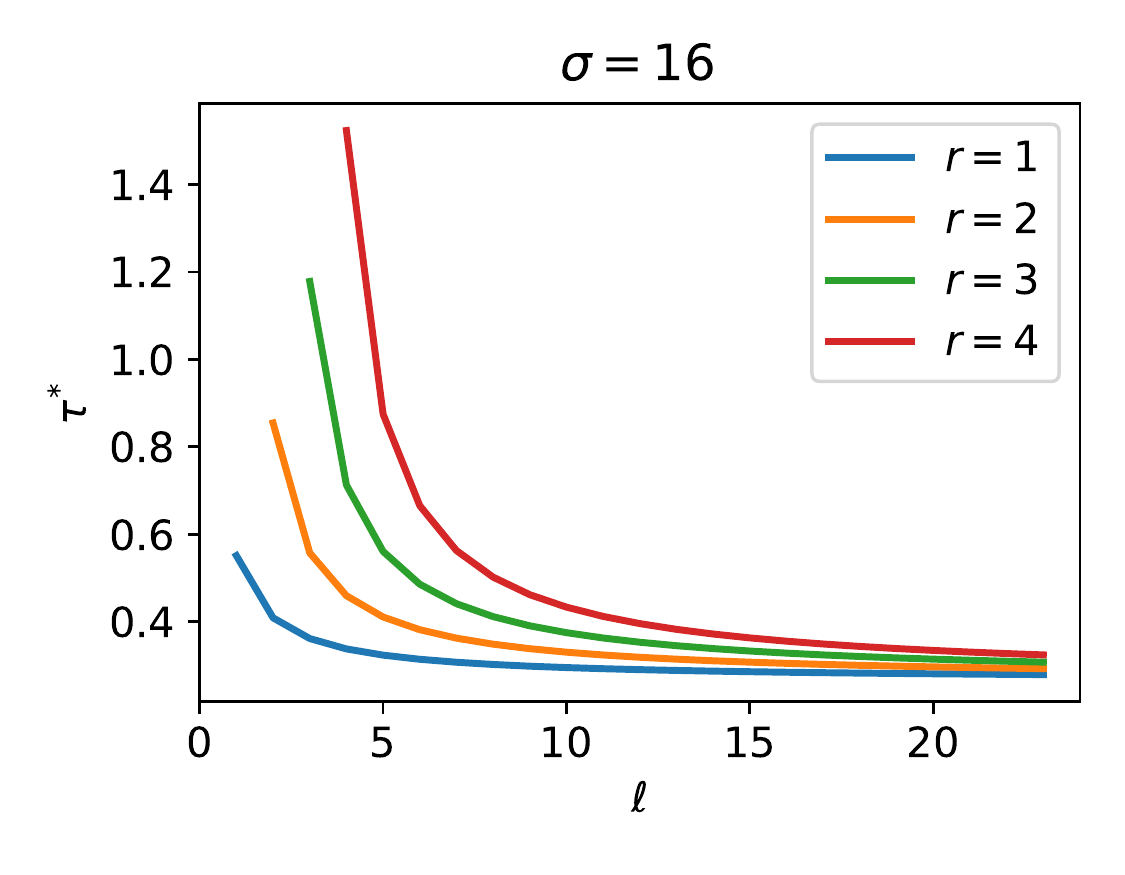}
\caption{Optimal thresholds $\tau^{*}$ for various parameters.}
\label{fig:tau}
\end{figure}

We fix the optimal threshold $\tau^{*}$ based on the search cost model.
After appending a new sketch to $L_v$, $\tau^{*}$ can be used to determine whether to split $v$ depending on $|L_v|$ to maintain the smaller search cost.
If $v$ is not split, then the search cost is $C_{\mathrm{leaf}}(v)$.
If $v$ is split into $k$ new leaves $u_1, u_2, \dots, u_k$, then the new search cost is
\begin{equation*}
C_{\mathrm{in}}(v) + \sum_{i=1}^{k} C_{\mathrm{leaf}}(u_i).
\end{equation*}
We assume that node $v$ is at level $\ell \geq r$.
Since the total length of $L_{u_1}, L_{u_2}, \dots, L_{u_k}$ is $|L_v|$, it holds that
\begin{equation*}
\sum_{i=1}^{k} C_{\mathrm{leaf}}(u_i) = P(\ell + 1) \times |L_v| \times \Ceil{\log_2 \sigma}.
\end{equation*}
Thus, splitting $v$ can maintain the smaller search cost if
\begin{equation}
\label{eq:tauopt}
|L_v| > \frac{P(\ell)}{P(\ell) - P(\ell + 1)} \times \frac{F_{\mathrm{in}}(\ell)}{\Ceil{\log_2 \sigma}} =: \tau^{*}.
\end{equation}
Given $r$ and $\sigma$, the optimal thresholds $\tau^{*}$ are determined for each level $\ell$ and pre-computable.
\fref{fig:tau} shows optimal thresholds $\tau^{*}$ for various parameters $r$ and $\sigma$.

\textbf{Exception Case.}
We need to address the exception when $\ell < r$, because the divisor of $\tau^{*}$ becomes zero, i.e., $P(\ell) = P(\ell + 1)$.
The occurrence of the exception is intuitively correct because the search always traverses all nodes at level $\ell \leq r$, and splitting a leaf at level $\ell < r$ just generates redundant nodes locally.

We fix $\tau^{*}$ to zero for $\ell < r$ since we cannot determine $\tau^{*}$ by \qref{eq:tauopt}.
Instead, we incrementally compute the total search cost of DyFT, defined as
\begin{equation*}
C_{\mathrm{trie}} = \sum_{v \in V_{\mathrm{in}}} C_{\mathrm{in}}(v) + \sum_{v \in V_{\mathrm{leaf}}} C_{\mathrm{leaf}}(v),
\end{equation*}
where $V_{\mathrm{in}}$ and $V_{\mathrm{leaf}}$ are sets of inner nodes and leaves in DyFT, respectively.
In the search phase, we compare the current cost $C_{\mathrm{trie}}$ with the computational cost of linear search for $X$, i.e., $C_{\mathrm{ls}} = n \times \Ceil{\log_2 \sigma}$.
If $C_{\mathrm{ls}} \leq C_{\mathrm{trie}}$, we perform linear search for $x_i \in X$ to avoid redundant node traversal;
otherwise, we perform \FuncFont{Search} in \gref{algo:asft}.
\gref{algo:asft_opt} shows the modified search algorithm.
The switching approach enables us to select the faster search algorithm depending on the configuration of DyFT.

\begin{algorithm}[tb]
\footnotesize

\DontPrintSemicolon

\SetKwFunction{KwFnSearchOpt}{\FuncFont{Search}$^{*}$}

\Fn{\KwFnSearchOpt{$y,r$}}{
\uIf(\tcp*[f]{Perform linear search}){$C_{\mathrm{ls}} \leq C_{\mathrm{trie}}$}{
    $R \gets \emptyset$\;
    \For{$x_i \in X$}{
        \lIf{$H(x_i,y) \leq r$}{
            Append $x_i$ to $R$
        }
    }
    \KwRet $R$\;
}
\Else(\tcp*[f]{Perform trie search in \gref{algo:asft}}){
    \KwRet $\FuncFont{Search}(y,r)$\;
}
}
\caption{Modified search algorithm of DyFT}
\label{algo:asft_opt}
\end{algorithm}

\textbf{Weighting Factor.}
In practice, the computational costs of $C_{\mathrm{in}}$ and $C_{\mathrm{leaf}}$ depend on the implementation of DyFT and the configuration of a computing machine.
To address the gap between the theoretical and practical costs, we introduce a weighting factor for inner nodes $W_{\mathrm{in}}$ and adjust the search cost for inner node $v$ by $W_{\mathrm{in}} \times C_{\mathrm{in}}(v)$.
We search a value of $W_{\mathrm{in}}$ that supports fast searches by using a synthetic dataset of random sketches generated from a uniform distribution.

\subsection{Complexities}
\label{sect:asft:complex}

We simply assume that $\tau$ is constant and derive the complexities shown in \tref{tab:review}.
The similarity search consists of traversing DyFT nodes, accessing posting lists and verifying candidates.
The number of traversed nodes is bounded by $O(m^{r+2})$ when assuming the complete $\sigma$-ary trie \cite{arslan2002dictionary}; thus, the traversal time is $O(m^{r+2})$.
The access time for each posting list is $O(1)$ because the length of each posting list is bounded by constant $\tau$.
Therefore, the search time complexity is $O(m^{r+2}) + V_\mathrm{dyft}$, where $V_\mathrm{dyft}$ is the verification time for the  obtained candidates.

Insertion is performed by traversing DyFT nodes in $O(m)$ time and splitting the posting list in $O(1)$ time.
Deletion is also performed by traversing DyFT nodes in $O(m)$ time and removing a leaf in $O(1)$ time.
Thus, the update time is $O(m)$.
The memory complexity is $O(mn)$ since the number of nodes is bounded by $O(mn)$.

\textbf{Multi-index Variant \ASFTP{}.}
The similarity search of DyFT is inefficient for large $r$ as the complexity is exponential to $r$.
We can relax the time using the multi-index approach \cite{greene1994multi}.
In the same manner as MIH, we define $q$ DyFT structures for each block.
We call this multi-index variant \emph{\ASFTP{}}.
The similarity search is performed on $q$ small DyFT structures with block length $m/q$ and threshold $\Floor{r/q}$.
The time complexity is $O(q (m/q)^{{r/q}+2}) +  V_\mathrm{dyft+}$, where $V_\mathrm{dyft+}$ is the verification time for the obtained candidates.
The update time and memory complexities are the same as those of DyFT.

\section{Node Implementation}
\label{sect:impl}

A node implementation is also significant to enhance the performance of DyFT.
This section presents \emph{modified adaptive radix tree (MART)}, which is an efficient node implementation for DyFT.
We first give observations for node implementations and then present our scheme of implementing MART.
Subsequently, we describe the data structure of MART.

\subsection{Observation and Implementation Scheme}
\label{sect:impl:obs}

We consider a data structure for an inner node that maps edge labels to child pointers.
A simple data structure referred to as the \emph{array form} is a pointer array of length $\sigma$ whose $c$-th element has the child pointer with edge label $c$.
The array form can directly obtain the child pointer for a given $c$.
Using the array form as a baseline, we provide the following observations for node implementations.

\textbf{Observation A.}
For binary sketches (i.e., $\sigma = 2$), the array form is memory-efficient because most inner nodes have two children and most elements of the array are used.
By chunking bits in binary sketches and suppressing the height of DyFT, we can reduce cache misses caused by node-to-node traversals and enhance time performance, as observed in prior studies \cite{leis2013adaptive,binna2018hot,boehm2011efficient}.

\textbf{Observation B.}
For integer sketches with large $\sigma$, inner nodes around the root have many children, but those around leaves have few children.
The array form is inefficient for nodes with few children because most elements of the array are empty.
Memory efficiency can be improved by introducing several data structures depending on the number of children, as suggested in prior studies \cite{leis2013adaptive,kanda2019b,zhang2018surf}.
Although {adaptive radix tree (ART)} \cite{leis2013adaptive} is a successful data structure in this approach, it was designed for byte edge labels and lacks generality to $\sigma$.

\textbf{Scheme.}
We assume $\sigma \leq 256$, following practical settings of similarity-preserving hashing techniques \cite{li2017linearized,li2010b,tabei2011sketchsort}.
MART reorganizes integer sketches into byte sketches to suppress DyFT's height (from Observation A) and represents DyFT nodes from byte sketches using a modified ART data structure (from Observation B).
\ssref{sect:impl:pack}{sect:impl:node} present the former and latter approaches, respectively.

\subsection{Byte Packing and Fast Computation}
\label{sect:impl:pack}

To efficiently handle integer sketches as byte sketches, we pack $z = \Floor{\log_{\sigma}(256)}$ integers $c_1,c_2,\dots,c_z$ into byte $b = \sum_{i = 1}^{z} c_i \sigma^{i-1} < \sigma^z \leq 256$.
In this manner, we convert an integer sketch $x = (c_1,c_2,\dots,c_m ) \in \Sigma^{m}$
into byte sketch $x' = (b_1,b_2,\dots,b_{m'})$ of length $m' = \Ceil{m/z}$.
In what follows, $H(b,b')$ denotes the Hamming distance between two integer sequences $c_1,c_2,\dots,c_z$ and $c'_1,c'_2,\dots,c'_z$ packed in two bytes $b$ and $b'$, respectively.

Through the packing, we build a DyFT structure from byte sketches and perform the similarity search using a given byte sketch.
When we visit an inner node $v$ during the search, we face the small problem corresponding to Lines \ref{line:asft:case1_beg}--\ref{line:asft:case2_end} in \gref{algo:asft}.
\begin{problem}%
\label{prob:asft}
Given an inner node $v$, byte label $b$, and radius $r'$, find children $u_1,u_2,\dots,u_k$ of $v$ with edge byte labels $a_1,a_2,\dots,a_k$ such that $H(a_i,b) \leq r'$.
\end{problem}

\begin{figure}[tb]
\centering
\includegraphics[scale=0.5]{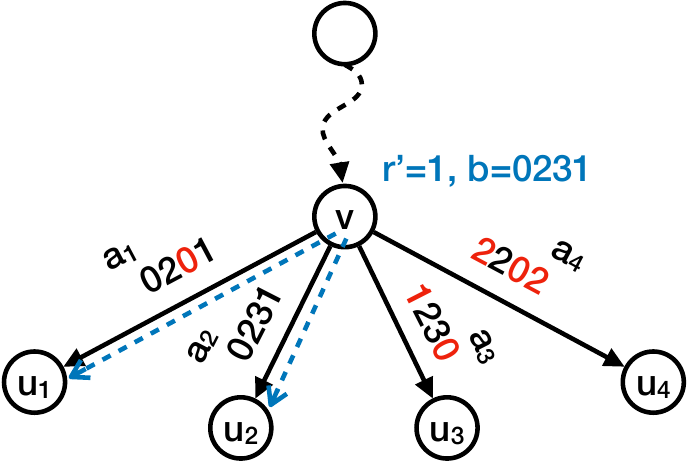}
\caption{
Example of Problem \ref{prob:asft} when $r' = 1$ and $\Sigma = \{0,1,2,3\}$.
The byte labels are denoted in unpacked format.
Children $u_1$ and $u_2$ are the solution because $H(a_1,b) = 1$ and $H(a_2,b) = 0$, while $u_3$ and $u_4$ are not the solution because $H(a_3,b) = 2$ and $H(a_4,b) = 3$.
}
\label{fig:prob1}
\end{figure}

\fref{fig:prob1} shows an example of Problem \ref{prob:asft}.
If $r' = 0$, we just look up a child with edge label $b$.
If $r' > 0$, we have the two approaches:
\LS{} visits all children of $v$ and computes the Hamming distances for the edge labels; 
\BF{} generates a set of all byte labels $A = \{a : H(a,b) \leq r'\}$ and looks up the children of $v$ with edge labels $a \in A$.
MART performs one of these approaches according to the configuration of a given inner node, as presented in \sref{sect:impl:node}.

To quickly perform the approaches without unpacking byte labels,
we introduce two $\sigma^z \times \sigma^z$ tables $\HDT$ and $\LUT$.
$\HDT$ is used in \LS{}, whose $b$-th row stores the Hamming distances between $b$ and all byte labels $a$, i.e., $\HDT[b,a] := H(a,b)$.
$\LUT$ is used in \BF{}, whose $b$-th row stores all byte labels $a$ sorted in ascending order of $H(a,b)$.
We can simply generate $A$ by scanning the elements of $\LUT[b,i]$ for $i = 0,1,\dots$ until we encounter $\HDT[b,\LUT[b,i]] > r'$.
Both $\HDT$ and $\LUT$ are implemented as simple tables of byte elements and can be precomputed.
Thus, $\HDT$ and $\LUT$ contribute to quickly solving Problem \ref{prob:asft} with only up to 128 KB of memory without unpacking byte labels.

\subsection{Adaptive Data Structure for Inner Nodes}
\label{sect:impl:node}

\begin{figure}[tb]
\centering
\includegraphics[scale=0.5]{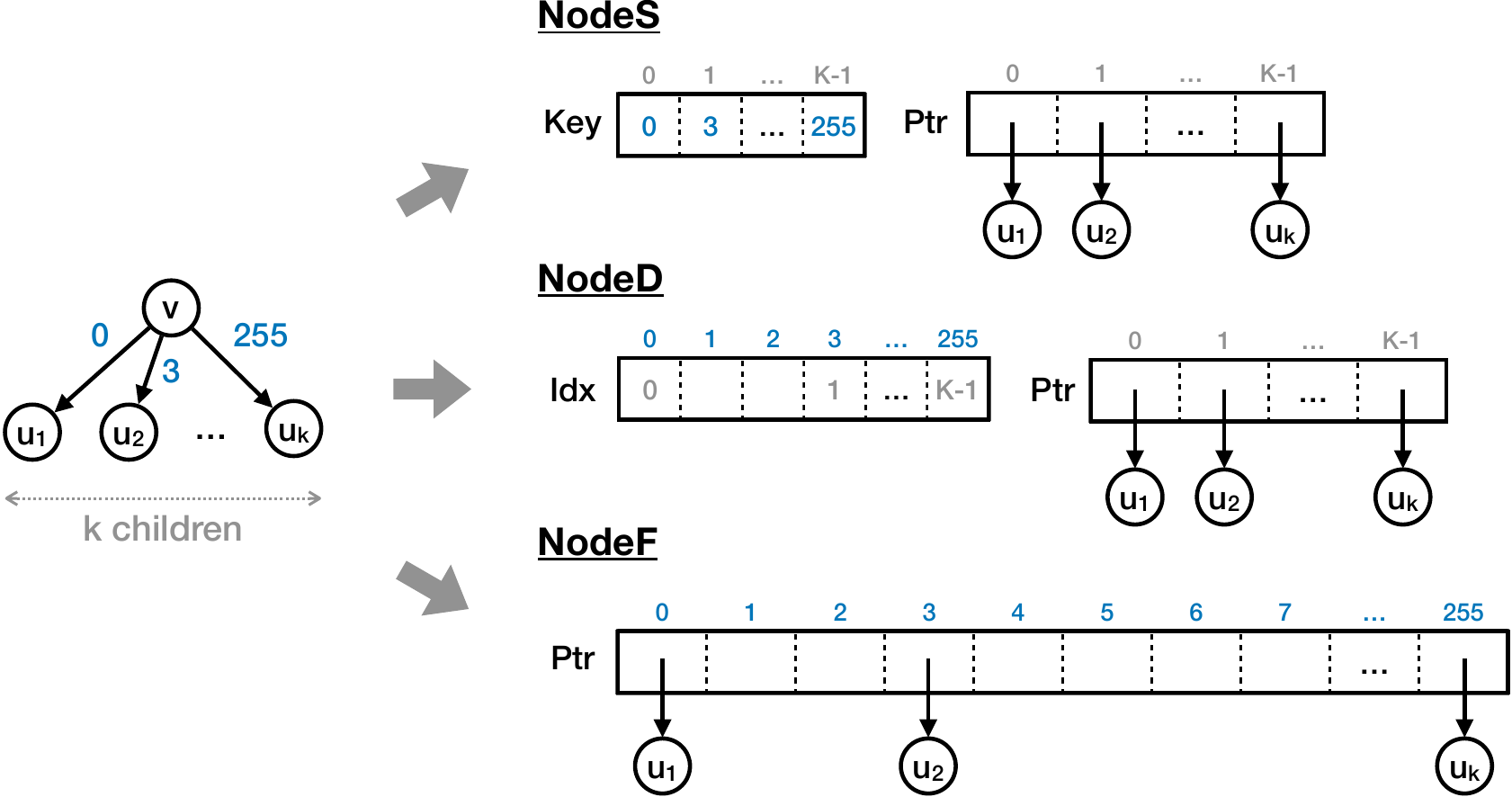}
\caption{
MART representations for node $v$ with $k$ children $u_1,u_2,\dots,u_k$.
The child pointer to $u_2$ with edge label 3 is stored in $\PtrS[1]$ such that $\KeyS[1] = 3$ in \NS{}, $\PtrD[\IdxD[3] = 1]$ in \ND{}, and $\PtrF[3]$ in \NF{}.
}
\label{fig:mart}
\end{figure}

\begin{algorithm}[tb]
\footnotesize

\DontPrintSemicolon

\SetKwFunction{KwFnS}{\FuncFont{NodeSearchS}}
\SetKwFunction{KwFnD}{\FuncFont{NodeSearchD}}
\SetKwFunction{KwFnF}{\FuncFont{NodeSearchF}}

\caption{MART search algorithms for Problem \ref{prob:asft}}
\label{algo:mart}

\Input{Inner node $v$, byte label $b$, and radius $r'$}
\Output{Set of child pointers $T$}
\BlankLine

\Fn(\tcp*[f]{\NS{}}){\KwFnS{$v,b,r'$}}{
$T \gets \emptyset$\;
\uIf(){$r' = 0$}{
    \tcp{Instead, SIMD search can be used as presented in \cite{leis2013adaptive}.}
    \For{$i = 0,1,\dots,v.k - 1$}{
        \If{$v.\KeyS[i] = b$}{
            Append $v.\PtrS[i]$ to $T$\;
            \textbf{break}\;
        }
    }
}
\Else(){
    \For(\tcp*[f]{\LS{}}){$i = 0,1,\dots,v.k - 1$}{
        \If{$\HDT[b,v.\KeyS[i]] \leq r'$}{
            Append $v.\PtrS[i]$ to $T$\;
        }
    }
}
\KwRet $T$\;
}
\BlankLine

\Fn(\tcp*[f]{\ND{}}){\KwFnD{$v,b,r'$}}{
$T \gets \emptyset$\;
\uIf(){$r' = 0$}{
    \If{$v.\IdxD[b] \neq K+1$}{
        Append $v.\PtrD[v.\IdxD[b]]$ to $T$\;
    }
}
\Else(){
    \For(\tcp*[f]{\BF{}}){$i = 0,1,\dots,\sigma^z - 1$}{
        \uIf{$\HDT[b, \LUT[b,i]] > r'$}{
            \textbf{break}\;
        }\ElseIf{$v.\IdxD[\LUT[b,i]] \neq K+1$}{
            Append $v.\PtrD[v.\IdxD[\LUT[b,i]]]$ to $T$\;
        }
    }
}
\KwRet $T$\;
}
\BlankLine

\Fn(\tcp*[f]{\NF{}}){\KwFnF{$v,b,r'$}}{
$T \gets \emptyset$\;
\uIf(){$r' = 0$}{
    \If{$v.\PtrF[b] \neq \NIL$}{
        Append $v.\PtrF[b]$ to $T$\;
    }
}
\Else(){
    \For(\tcp*[f]{\BF{}}){$i = 0,1,\dots,\sigma^z - 1$}{
        \uIf{$\HDT[b, \LUT[b,i]] > r'$}{
            \textbf{break}\;
        }\ElseIf{$v.\PtrF[\LUT[b,i]] \neq \NIL$}{
            Append $v.\PtrF[\LUT[b,i]]$ to $T$\;
        }
    }
}
\KwRet $T$\;
}

\end{algorithm}

Although ART \cite{leis2013adaptive} is a space-efficient data structure for representing inner nodes with byte labels, the design is for standard trie structures and is redundant for DyFT.
For example, the path-compression technique of ART is not necessary for DyFT.
MART simply modifies ART and represents inner nodes of DyFT.
MART uses the following three types of data structures depending on the number of children.
Let us consider representing an inner node $v$ with $k$ children.
The three types of data structures are illustrated in \fref{fig:mart}, and their algorithms to Problem \ref{prob:asft} are presented in \gref{algo:mart}.

\textbf{\NS{} (NodeSparse)} is a data structure for storing node $v$ with $k$ children of no more than $K$, where $K$ is a constant parameter.
It consists of two arrays $\KeyS$ and $\PtrS$.
$\KeyS$ is a byte array of length $K$ that stores edge labels from $v$.
$\PtrS$ is a pointer array of length $K$ such that $\PtrS[i]$ stores the child pointer with edge label $\KeyS[i]$.
We maintain the arrays such that the first $k$ elements are used.
Problem \ref{prob:asft} is solved by performing \LS{} for the first $k$ elements of $\KeyS$.
If $r' = 0$, modern CPUs can quickly search the elements using SIMD instructions in parallel, as presented in \cite{leis2013adaptive}.
\FuncFont{NodeSearchS} shows the algorithm.

\textbf{\ND{} (NodeDense)} is a data structure for storing node $v$ with $k$ children no more than $K$.
It consists of two arrays $\IdxD$ and $\PtrD$.
$\IdxD$ is a byte array of length 256 to indicate positions of $\PtrD$.
$\PtrD$ is a pointer array of length $K$ such that $\PtrD[\IdxD[b]]$ stores the child pointer with edge label $b$.
$\IdxD[b] = K+1$ indicates that there is not a child pointer with $b$.
Problem \ref{prob:asft} is solved by performing \BF{} for $\IdxD$.
\FuncFont{NodeSearchD} shows the algorithm.

\textbf{\NF{} (NodeFull)} is a data structure for very large $k$ and consists of pointer array $\PtrF$ of length 256 such that $\PtrF[b]$ stores the child pointer with edge label $b$.
The data structure is identical to the array form.
Problem \ref{prob:asft} is solved by performing \BF{} for $\PtrF$.
\FuncFont{NodeSearchF} shows the algorithm.

Every data structure has a header of one byte to store the value of $k$.
Let $w$ be the word size in bits such as 32 or 64 bits.
\NS{} consumes $8 + 8K + wK$ bits, \ND{} consumes $8 + 8 \cdot 256 + wK$ bits, and \NF{} consumes $8 + 256w$ bits.
\NS{} is the most memory-efficient but uses \LS{} taking $O(k)$ time.
\ND{} is more memory-efficient than \NF{} when $K < 256 - 2048/w$.

With respect to time and space, \NS{} is efficient for small $K$, and \ND{} is efficient for large $K$.
We define \NS{} with $K = 2,4,8,16,32$ and \ND{} with $K = 64,128$.
An inner node with $k$ children of no more than 128 is represented in \NS{} or \ND{} such that $K$ is the smallest and no less than $k$.
An inner node with $k$ children of more than 128 is represented in \NF{}.
This adaptive selection allows child pointers to be stored space-efficiently.

\subsection{Compact Implementation for Leaves}
\label{sect:impl:leaf}

Finally, we briefly present a compact implementation of leaves.
Each leaf is represented as a pointer to the posting list.
We compress the pointers using a sparse direct address table \cite{norouzi2014fast} that groups $g$ pointers by concatenating the $g$ posting lists and reduces the number of pointers by a factor of $g$.
Given a leaf, the sparse direct address table can access the corresponding posting list using the identifier in $O(g/w)$ time.
DyFT sets $g = w$ to perform the access in constant time.
The implementation details are presented in \cite[Sect. 6]{norouzi2014fast}.

\section{Experiments}
\label{sect:ex}

\newcommand{\ChartWidth}{42mm}

\newcommand{\FastText}{\FuncFont{Text1M}}
\newcommand{\MNIST}{\FuncFont{MNIST8M}}
\newcommand{\Review}{\FuncFont{Review13M}}
\newcommand{\CP}{\FuncFont{CP216M}}

We evaluated the performances of DyFT and \ASFTP{} using three real-world vector datasets.
\FastText{} consists of 999,994 pre-trained continuous word vectors from English Wikipedia 2017 using fastText \cite{mikolov2018advances}, where each vector is a real number vector of 300 dimensions.
\Review{} consists of 12,886,488 book reviews in English from Amazon \cite{mcauley2013hidden}.
Each review is represented as a 9,253,464-dimensional binary fingerprint of which each dimension represents the presence or absence of a word.
\CP{} consists of 216,121,626 compound-protein pairs in the STITCH database \cite{kuhn2009stitch}, where each pair is represented as a 3,621,623-dimensional binary fingerprint.

We converted real number vectors in \FastText{} into binary sketches using Charikar's simhash algorithm \cite{charikar2002similarity} and integer sketches using the GCWS algorithm \cite{li2017linearized}.
We converted binary vectors in \Review{} and \CP{} into binary or integer sketches using Li's mihhash algorithm \cite{li2010b}.

We constructed an index of similarity search methods by inserting sketches in a dataset in random order.
We measured the elapsed insertion time and required memory usage for the construction.
We produced a query set by randomly sampling 1,000 sketches from each dataset and measured the average similarity search time per query.

We evaluated $\sigma = 16$ for integer sketches following the practical considerations in \cite{li2017linearized,li2010b}.
We evaluated DyFT and HWT (without the multi-index approach) using short sketches of $m = 32$ and small radii $r \leq 4$.
We evaluated \ASFTP{}, MIH, HmSearch, and GV (with the multi-index approach) using long sketches of $m = 64$ and large radii $r \leq 10$.
We fixed $W_{\mathrm{in}} = 0.5$ based on experiments using a dataset of 10 million random sketches.

We conducted all experiments on one core of quad-core Intel Xeon CPU E5--2680 v2 clocked at 2.8 GHz in a machine with 256 GB of RAM running the 64-bit version of CentOS 6.10 based on Linux 2.6.
We implemented all data structures in {C++17} and compiled source codes using g++ version 7.3.0 with optimization flags {-O3} and {-march=native}.
The code used in our experiments is available at \url{https://github.com/kampersanda/dyft}.

\subsection{Analysis for Optimal Threshold $\tau^{*}$}

\begin{figure}[tb]
\centering
\includegraphics[width=\ChartWidth]{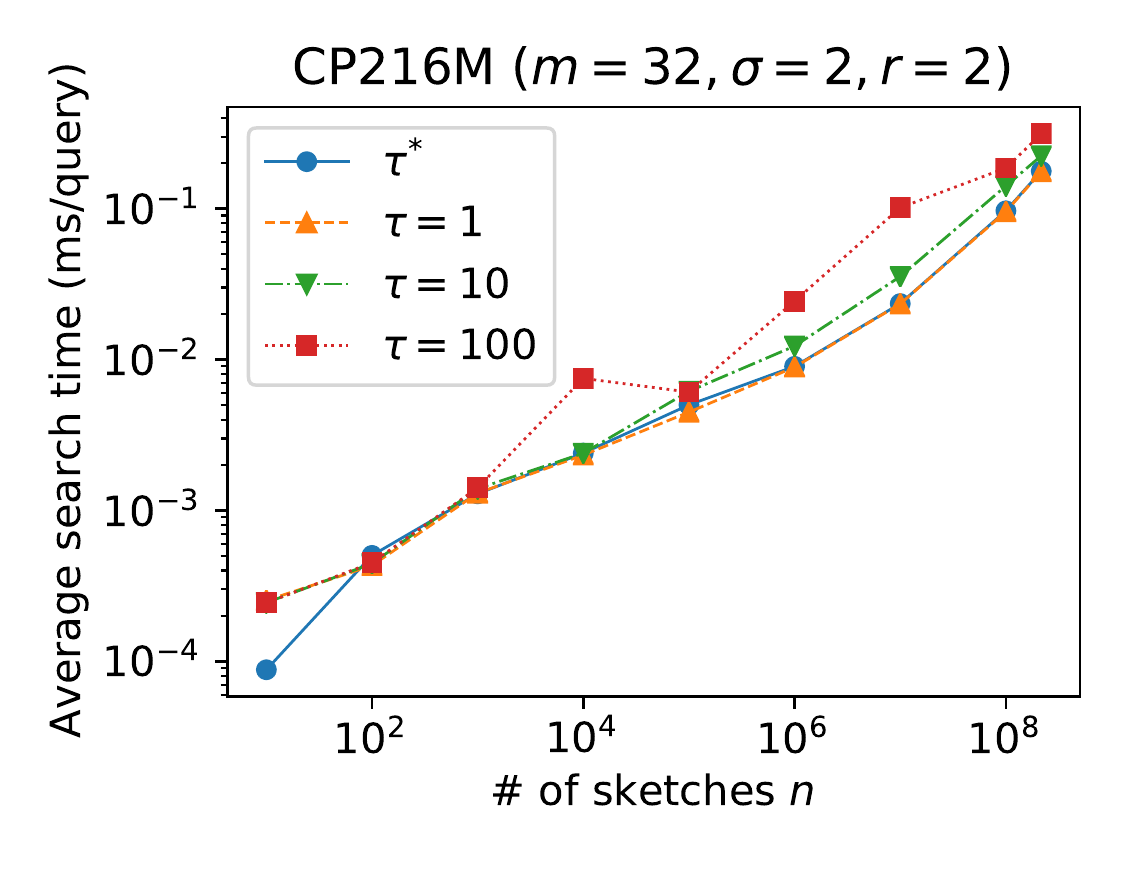}
\includegraphics[width=\ChartWidth]{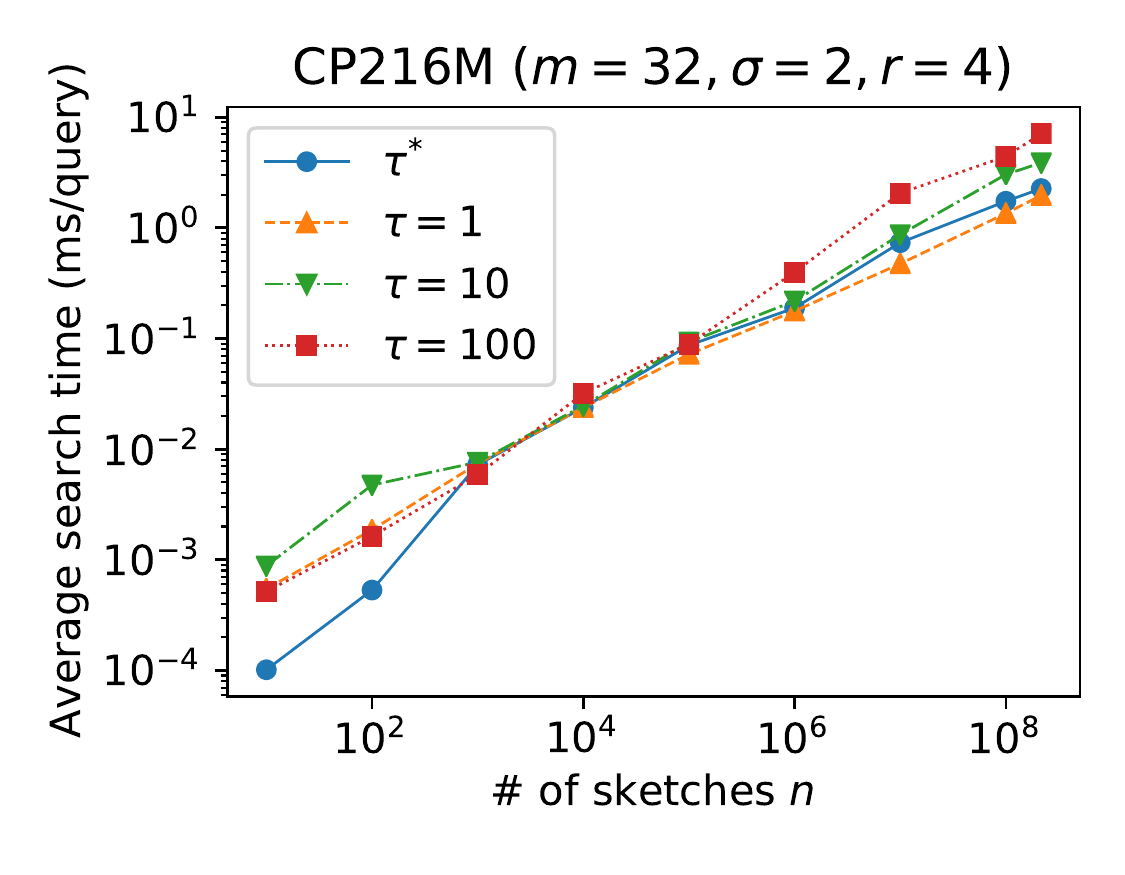}\\
\includegraphics[width=\ChartWidth]{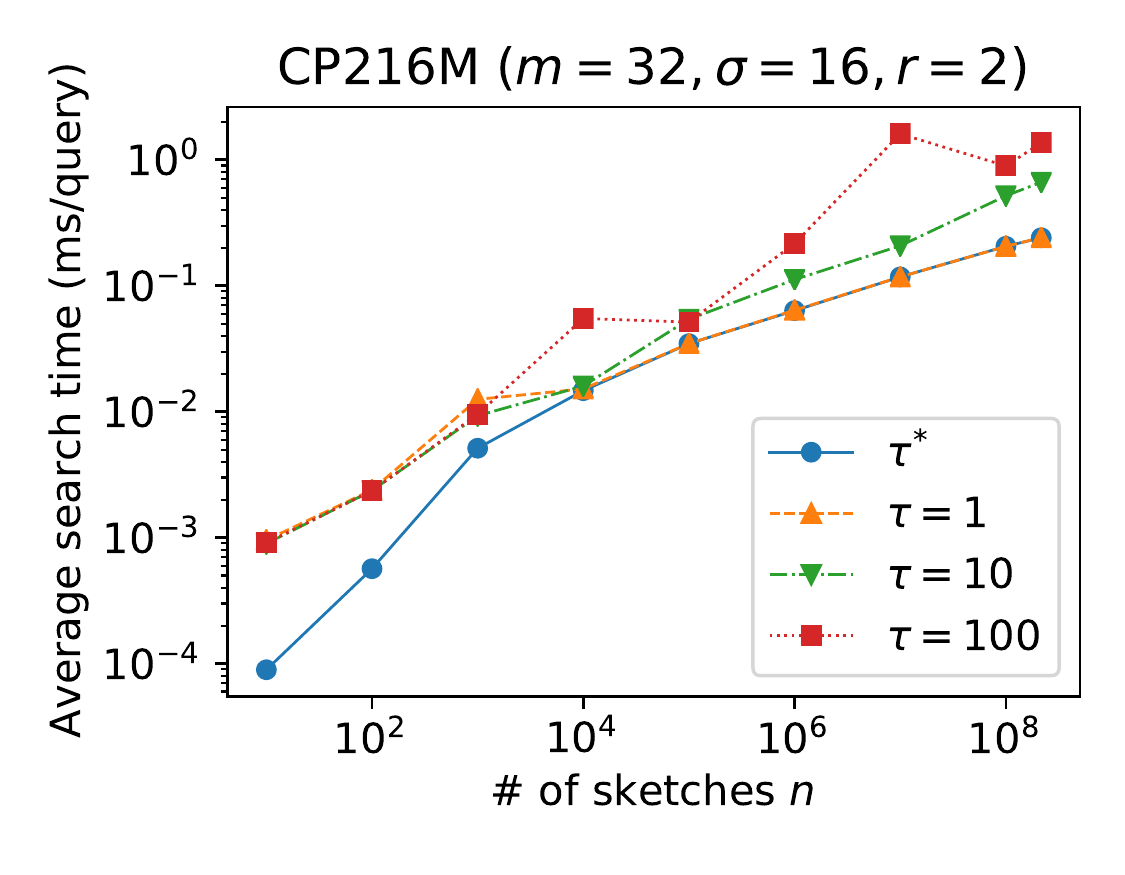}
\includegraphics[width=\ChartWidth]{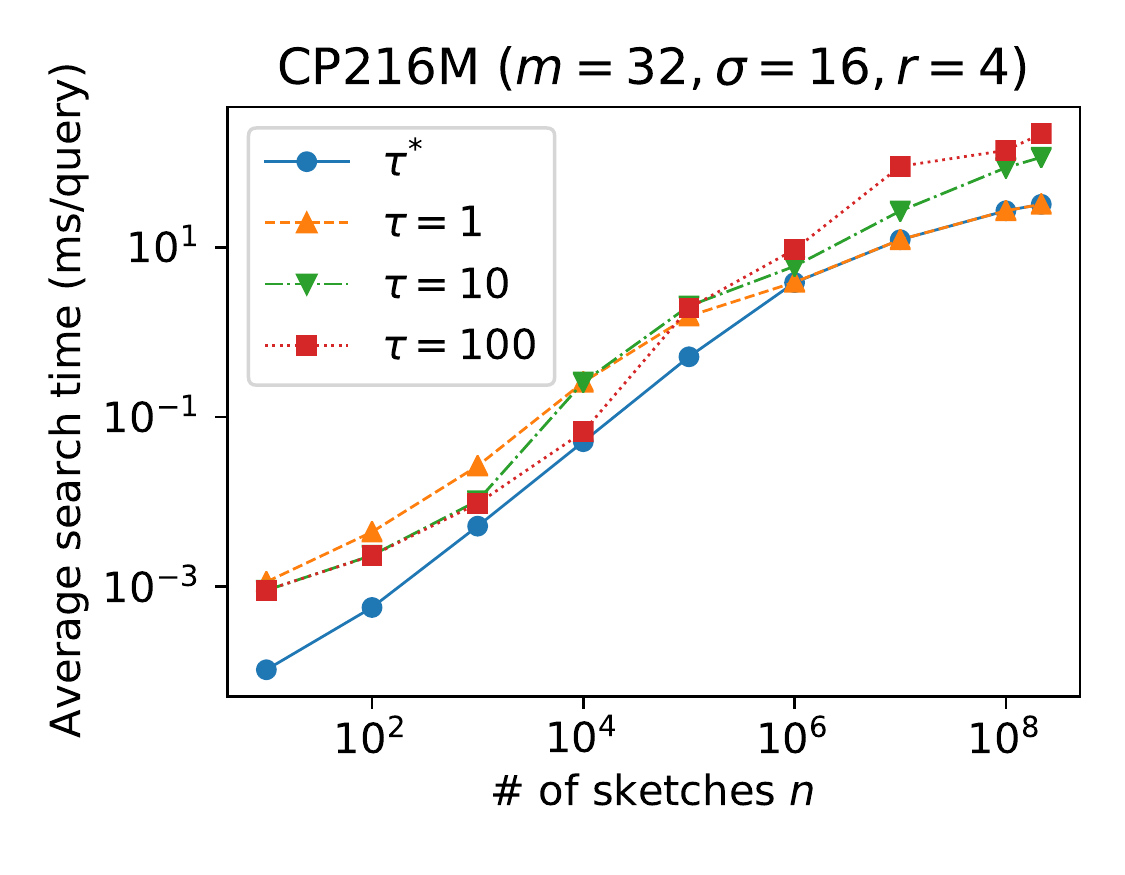}
\caption{
Results for optimal threshold $\tau^*$ and fixed thresholds $\tau = 1,10,100$ on \CP{}.
The charts show average search time in milliseconds for varying the number of sketches $n$ plotted in \textbf{logarithmic scale}.
}
\label{fig:splitthrs}
\end{figure}

We analyzed DyFT's performance with optimal threshold $\tau^{*}$ and fixed thresholds $\tau = 1, 10, 100$.
\fref{fig:splitthrs} shows the results of search time on \CP{} when $r = 2,4$.
The search time with $\tau^{*}$ was the fastest in most cases.
The effectiveness of $\tau^{*}$ could be observed especially when $\sigma = 16$ and $r=4$.
The search times with $\tau$ were reversed according to $n$, i.e., setting $\tau = 1$ provided faster searches for large $n$ while setting $\tau = 100$ provided faster searches for small $n$.
This demonstrated that $\tau$ is not efficient in dynamic settings where $n$ is varied.
On the other hand, $\tau^{*}$ maintained the fastest similarity search speed even when $n$ was varied.

\subsection{Analysis for Node Implementations}

We compared the performances of MART, the array form (Array), and the original ART \cite{leis2013adaptive}.
We evaluated each data structure when implementing inner nodes of DyFT.
Both Array and ART did not apply the byte-packing technique.
The aim of the comparison with ART is to observe the effectiveness of the byte-packing technique;
hence, we did not implement unnecessary techniques of ART such as path compression.

\begin{table}[tb]
\centering
\caption{Results for Node Implementations on \Review{} ($m=32$)}
\label{tab:mart}
\begin{tabular}{lrrrrrr} 
\toprule
 & \multicolumn{3}{c}{$\sigma = 2$ (binary)} & \multicolumn{3}{c}{$\sigma = 16$ (integer)} \\
\cmidrule(lr){2-4}
\cmidrule(lr){5-7}
$r$ & Array & ART & MART & Array & ART & MART \\
\midrule
 & \multicolumn{6}{c}{Search Time (ms) per Query} \\
\cmidrule(lr){3-6}
1 & 0.014 & 0.019 & \textbf{0.003} & \textbf{0.008} & 0.017 & \textbf{0.008} \\
2 & 0.12 & 0.17 & \textbf{0.02} & 0.18 & 0.36 & \textbf{0.16} \\
3 & 0.72 & 1.03 & \textbf{0.11} & 3.0 & 5.3 & \textbf{2.3} \\
4 & 3.5 & 4.8 & \textbf{0.8} & 32 & 48 & \textbf{21} \\
\midrule
 & \multicolumn{6}{c}{Insertion Time (sec)} \\
\cmidrule(lr){3-6}
1 & 16 & 20 & \textbf{7} & \textbf{16} & 20 & 20 \\
2 & 16 & 20 & \textbf{8} & \textbf{16} & 20 & 20 \\
3 & 16 & 20 & \textbf{8} & \textbf{16} & 20 & 20 \\
4 & 16 & 20 & \textbf{7} & \textbf{16} & 20 & 20 \\
\midrule
 & \multicolumn{6}{c}{Memory Usage (MB)} \\
\cmidrule(lr){3-6}
1 & \textbf{196} & 379 & 246 & 882 & 334 & \textbf{333} \\
2 & \textbf{196} & 379 & 249 & 882 & \textbf{335} & \textbf{335} \\
3 & \textbf{195} & 378 & 244 & 881 & \textbf{333} & 334 \\
4 & \textbf{184} & 350 & 202 & 880 & \textbf{335} & \textbf{335} \\
\bottomrule
\end{tabular}
\end{table}

\tref{tab:mart} shows the results of search time, insertion time, and memory usage on \Review{}.
They demonstrated the validity of our observations in \sref{sect:impl:obs}.
The search time of MART was the fastest in all cases.
Compared to Array, MART was at most \Times{6.3} faster for binary sketches and at most \Times{1.5} faster for integer sketches.
This suggests that suppressing DyFT's height with the byte-packing technique provides fast retrieval on Observation A.
Similarly, the insertion time of MART was the fastest for binary sketches due to the byte-packing technique, although Array was the fastest for integer sketches due to the simplest data structure.
With respect to memory usage, Array was the smallest for binary sketches but largest for integer sketches on Observations A and B;
ART and MART were the smallest for integer sketches on Observation B.
Overall, MART achieved relevant space-time trade-offs for both binary and integer sketches.

\subsection{Analysis for DyFT on Binary Sketches}

We compared the performances of DyFT and HWT.
HWT is the state-of-the-art method designed for dynamic similarity searches on binary sketches \cite{eghbali2019online}.
We implemented HWT using the original source code available at \url{https://github.com/sepehr3pehr/hwt}.

\begin{figure}[tb]
\centering
\includegraphics[width=\ChartWidth]{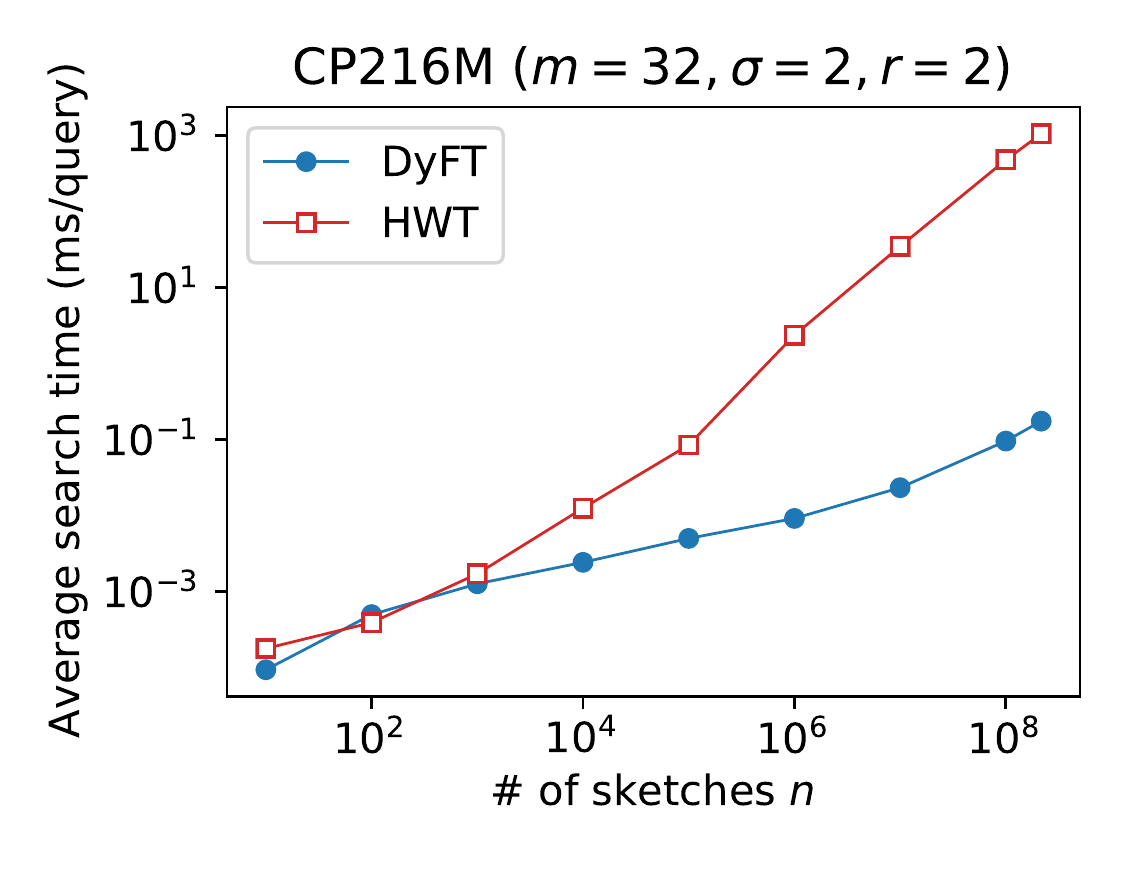}
\includegraphics[width=\ChartWidth]{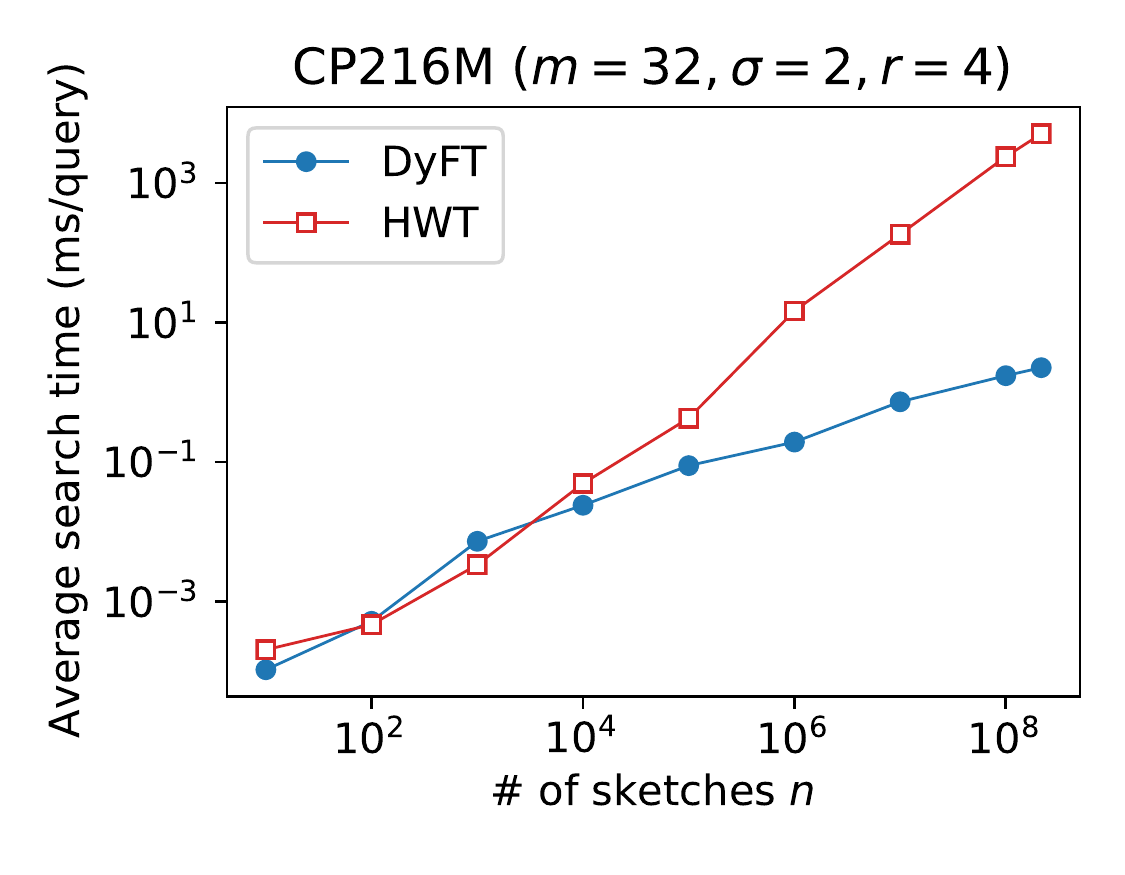}\\
\includegraphics[width=\ChartWidth]{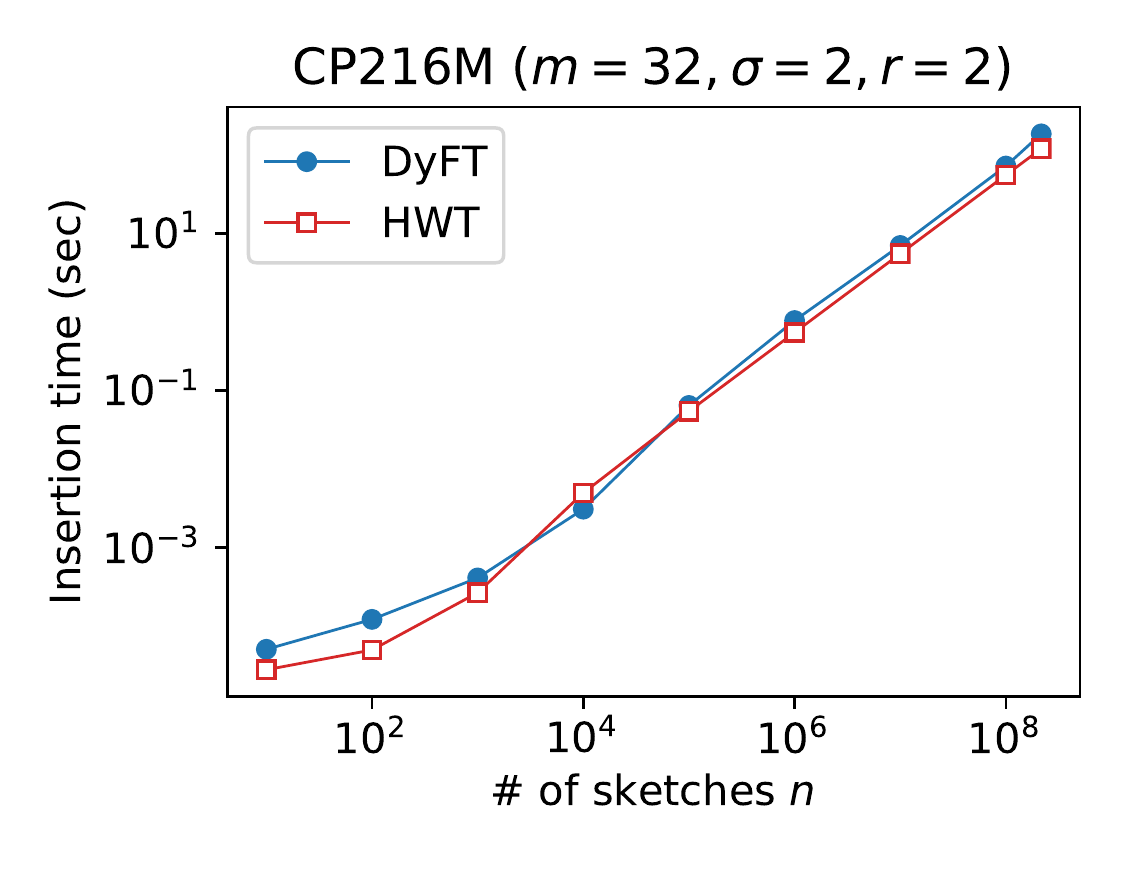}
\includegraphics[width=\ChartWidth]{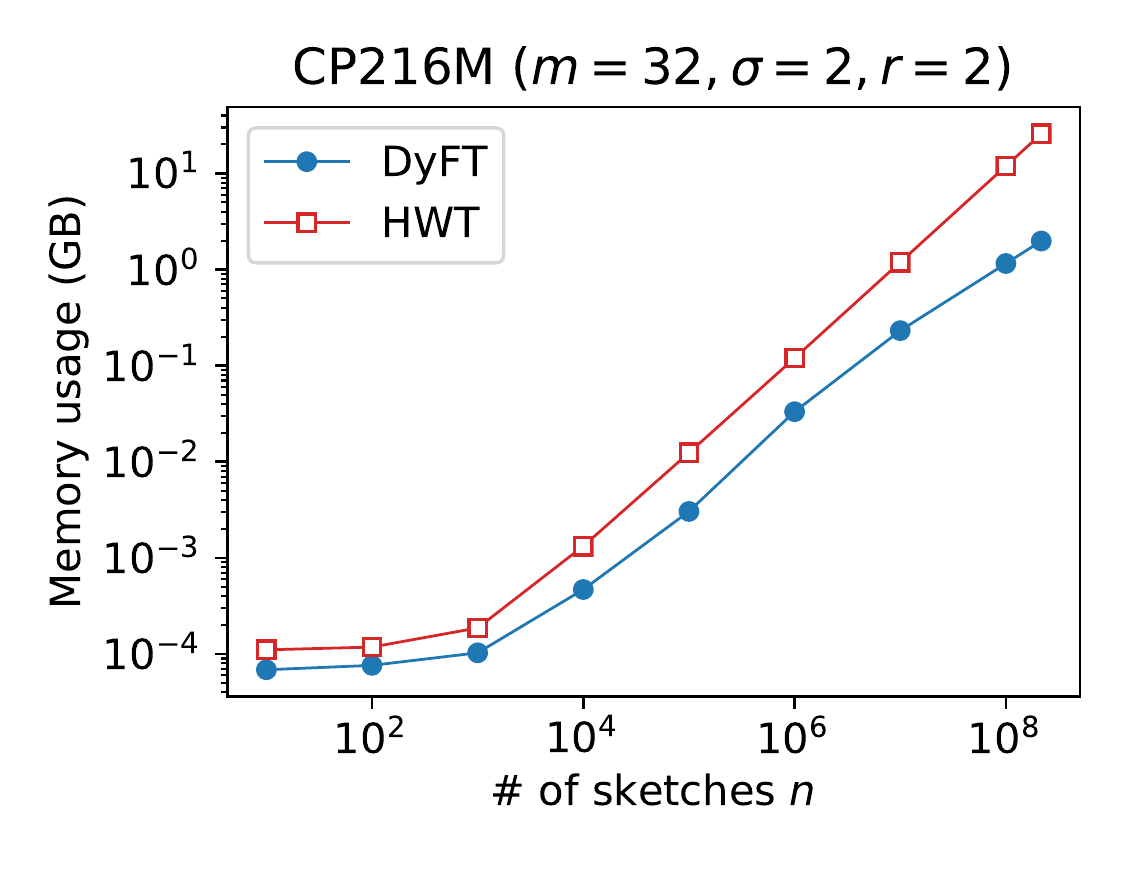}
\caption{
Results for DyFT and HWT on \CP{}.
The upper charts show average search time in milliseconds for varying the number of sketches $n$ when $r=2,4$.
The bottom-left chart shows  insertion time in minutes for varying $n$.
The bottom-right chart shows  memory usage in GB for varying $n$.
They are plotted in \textbf{logarithmic scale}.
}
\label{fig:asft_vs_hwt}
\end{figure}

\fref{fig:asft_vs_hwt} shows the results of search time, insertion time, and memory usage on \CP{}.
As $n$ increased, the search time of DyFT became faster than that of HWT.
This result is consistent with the search time complexities of DyFT and HWT, as HWT's complexity contains the factor of $O(\log n)$.
When $r=2$, DyFT was at most \Times{6000} faster than HWT.
Although HWT's insertion time complexity $O(m \log m)$ is worse than DyFT's complexity $O(m)$, the measured insertion times were not much different because $m$ was not significant.
Although the memory complexities of DyFT and HWT are the same, DyFT was at most \Times{13} more memory-efficient than HWT because of the node-omitting approach and MART.

\subsection{Analysis for \ASFTP{} on Binary Sketches}

\begin{figure*}[tb]
\centering
\includegraphics[width=\ChartWidth]{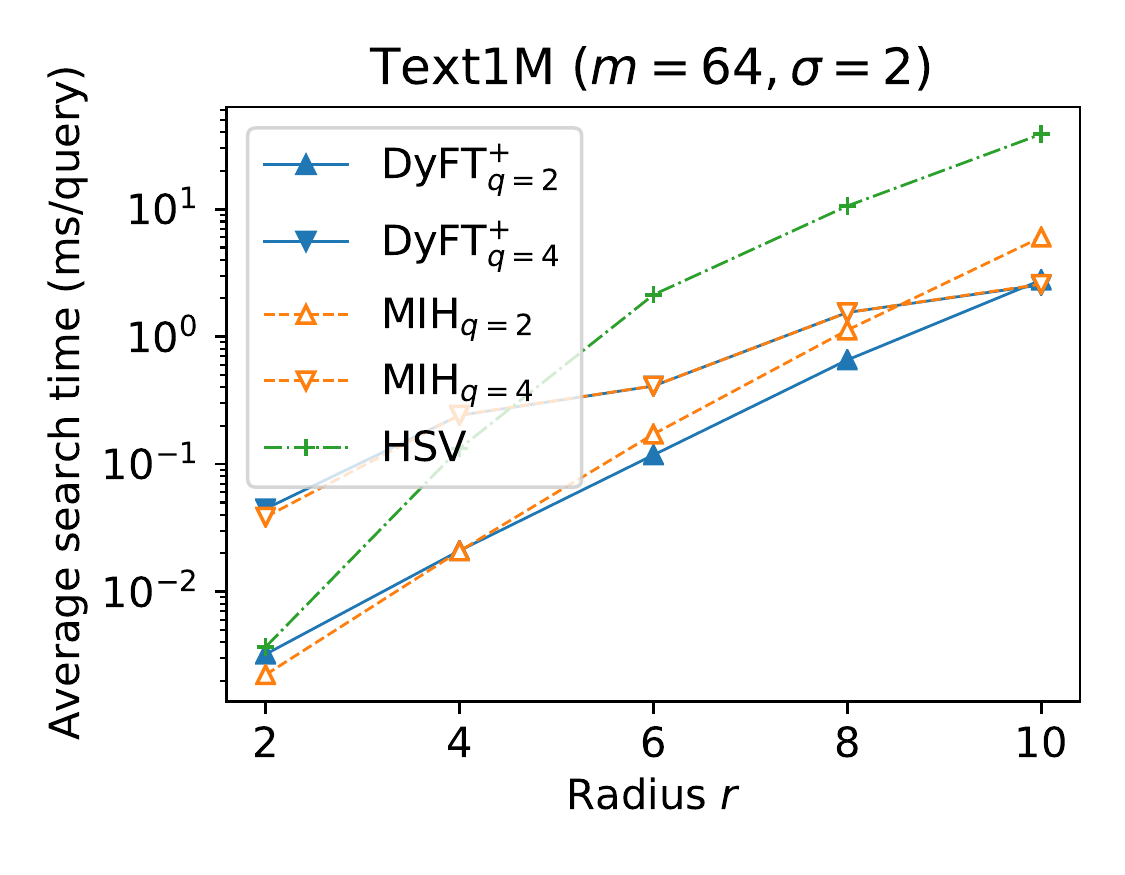}
\includegraphics[width=\ChartWidth]{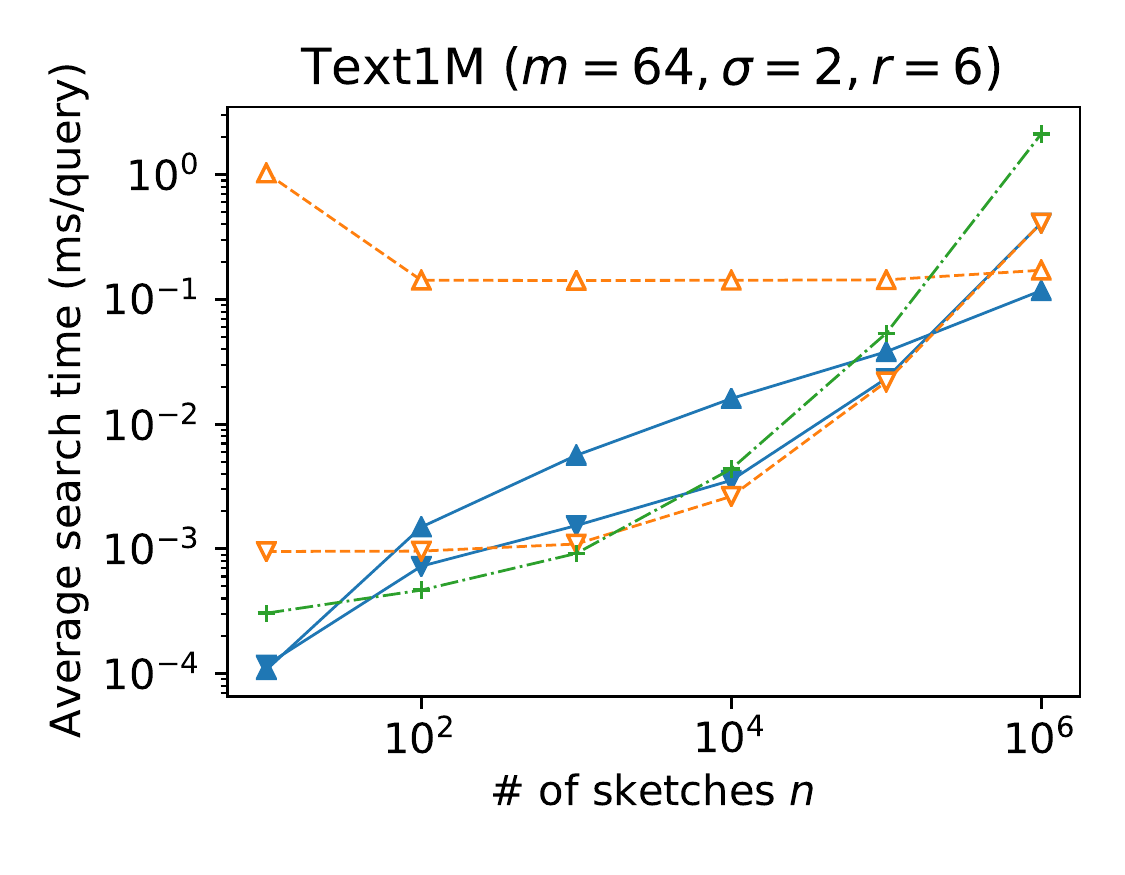}
\includegraphics[width=\ChartWidth]{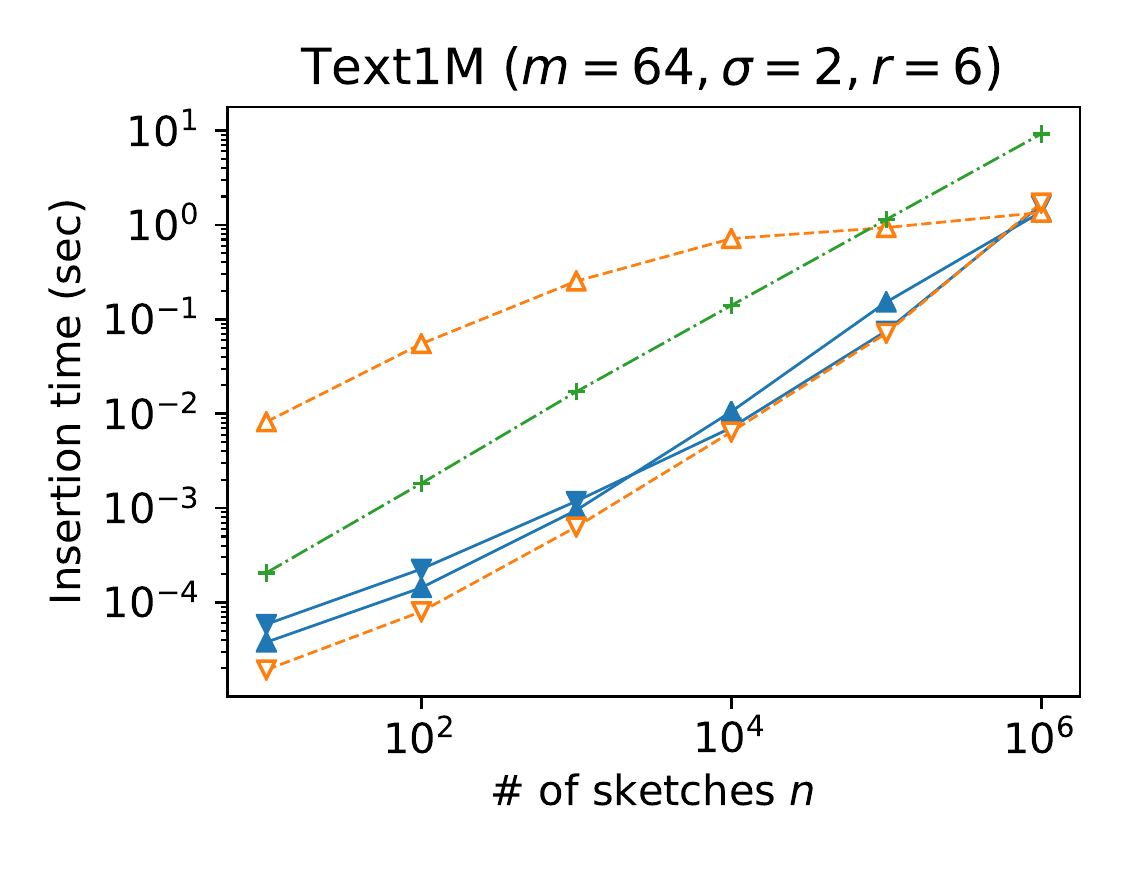}
\includegraphics[width=\ChartWidth]{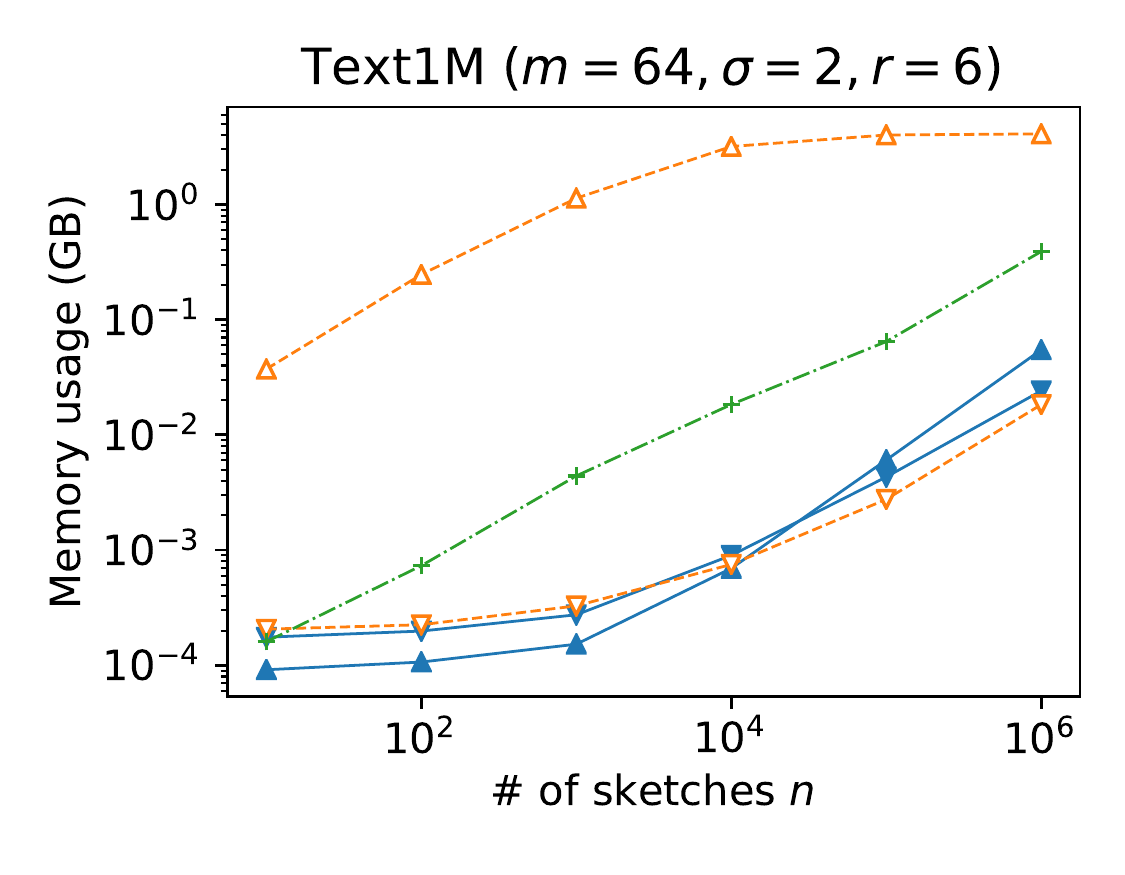}\\
\includegraphics[width=\ChartWidth]{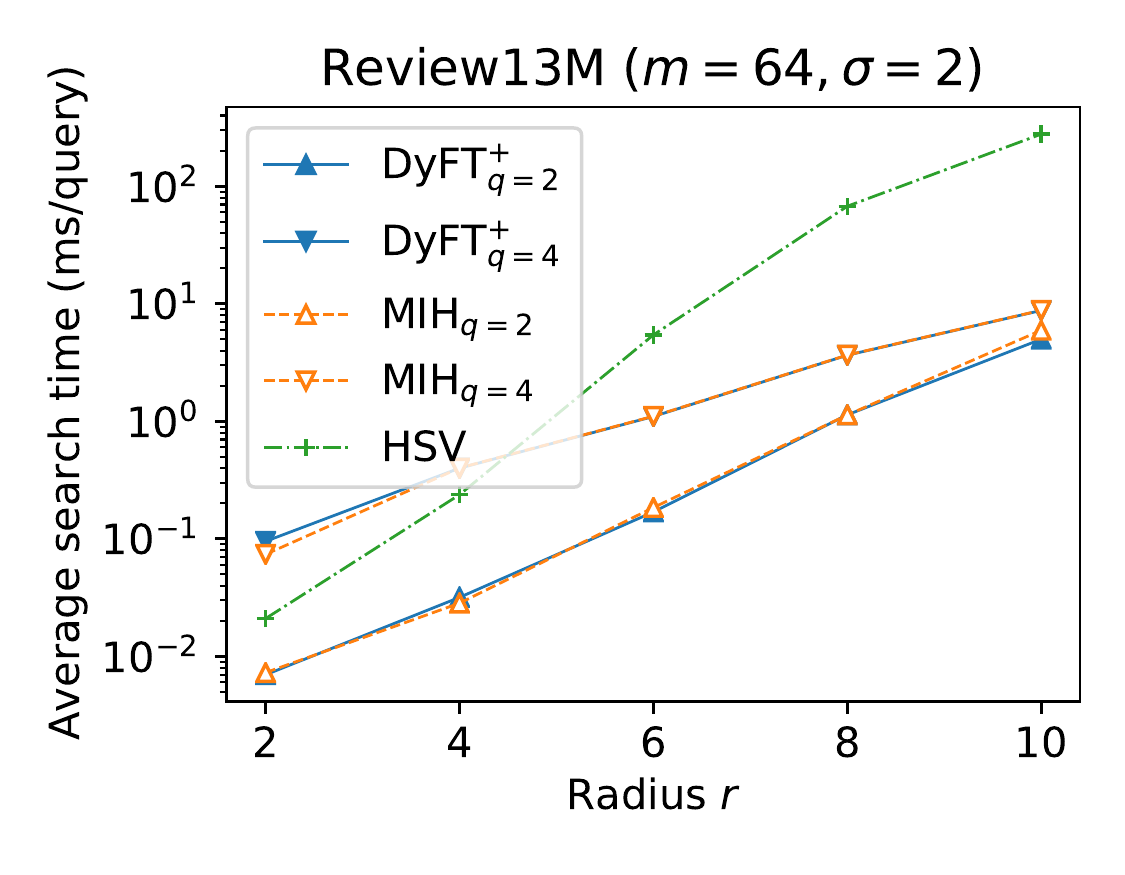}
\includegraphics[width=\ChartWidth]{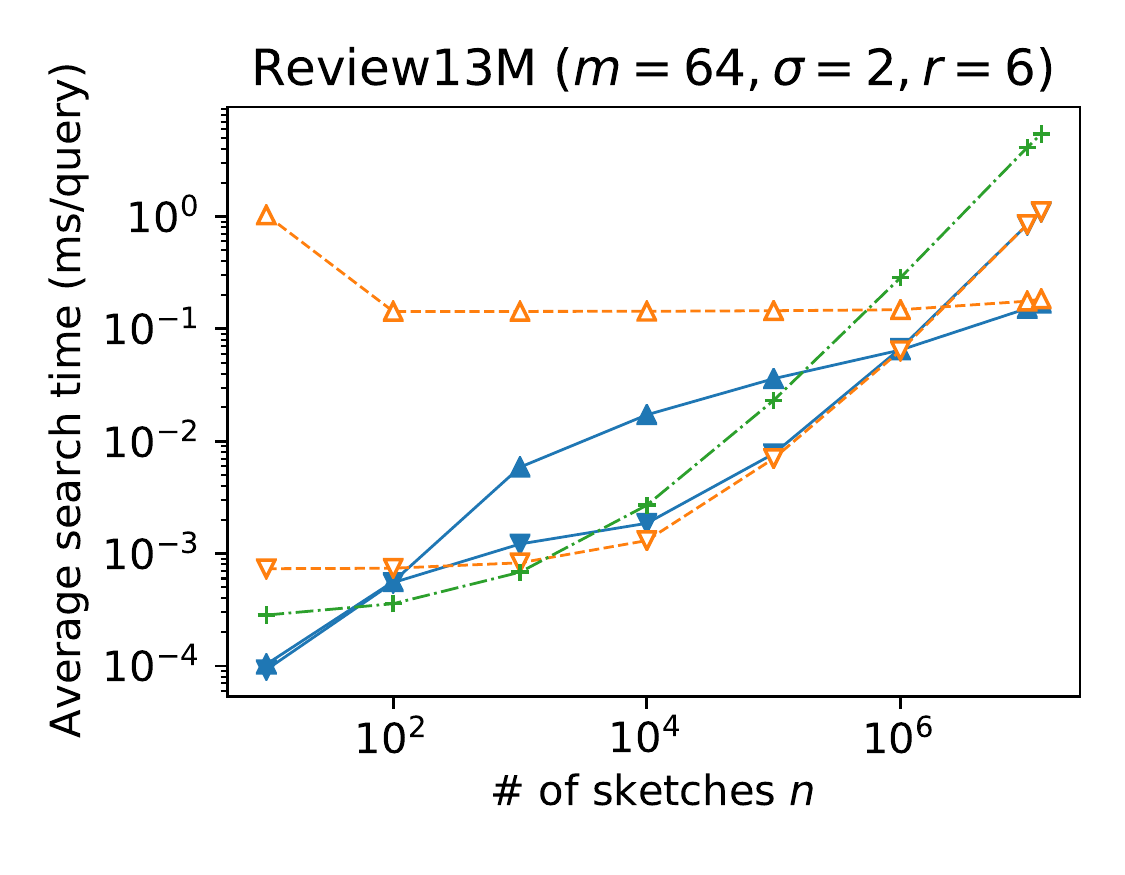}
\includegraphics[width=\ChartWidth]{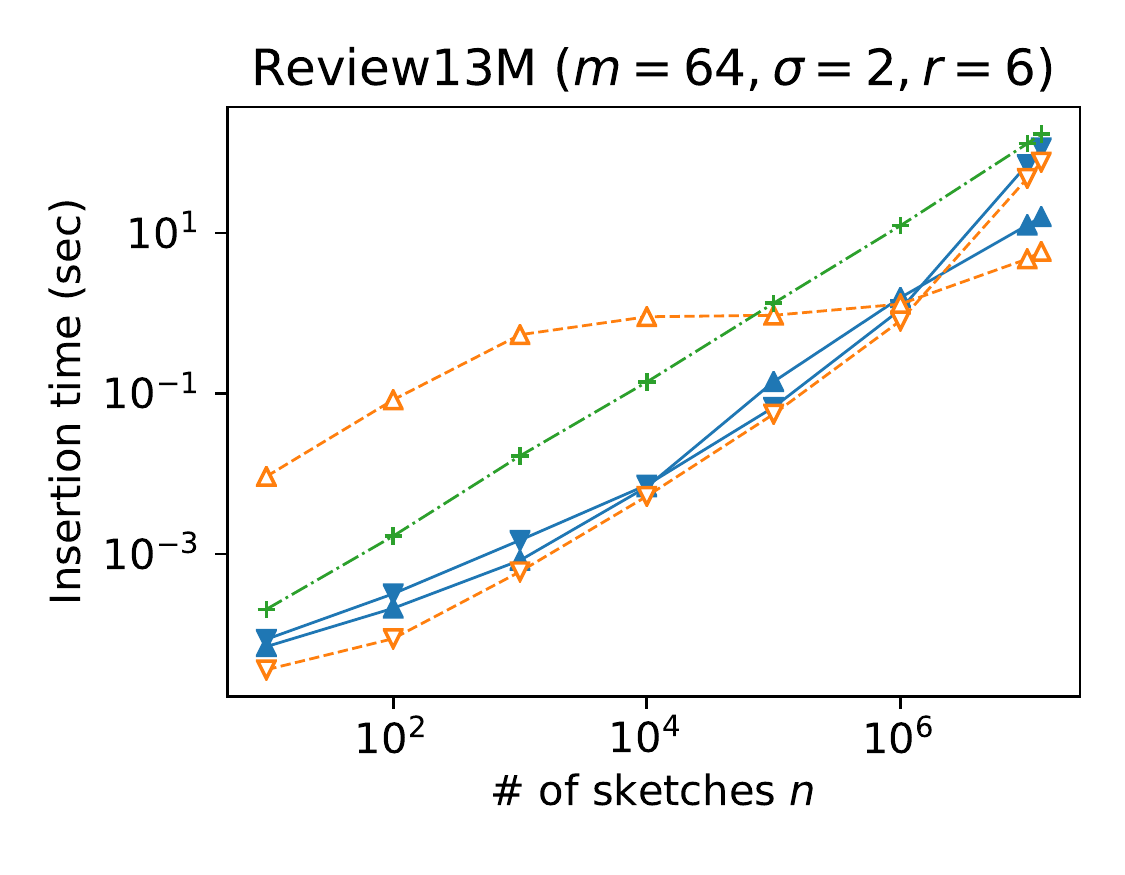}
\includegraphics[width=\ChartWidth]{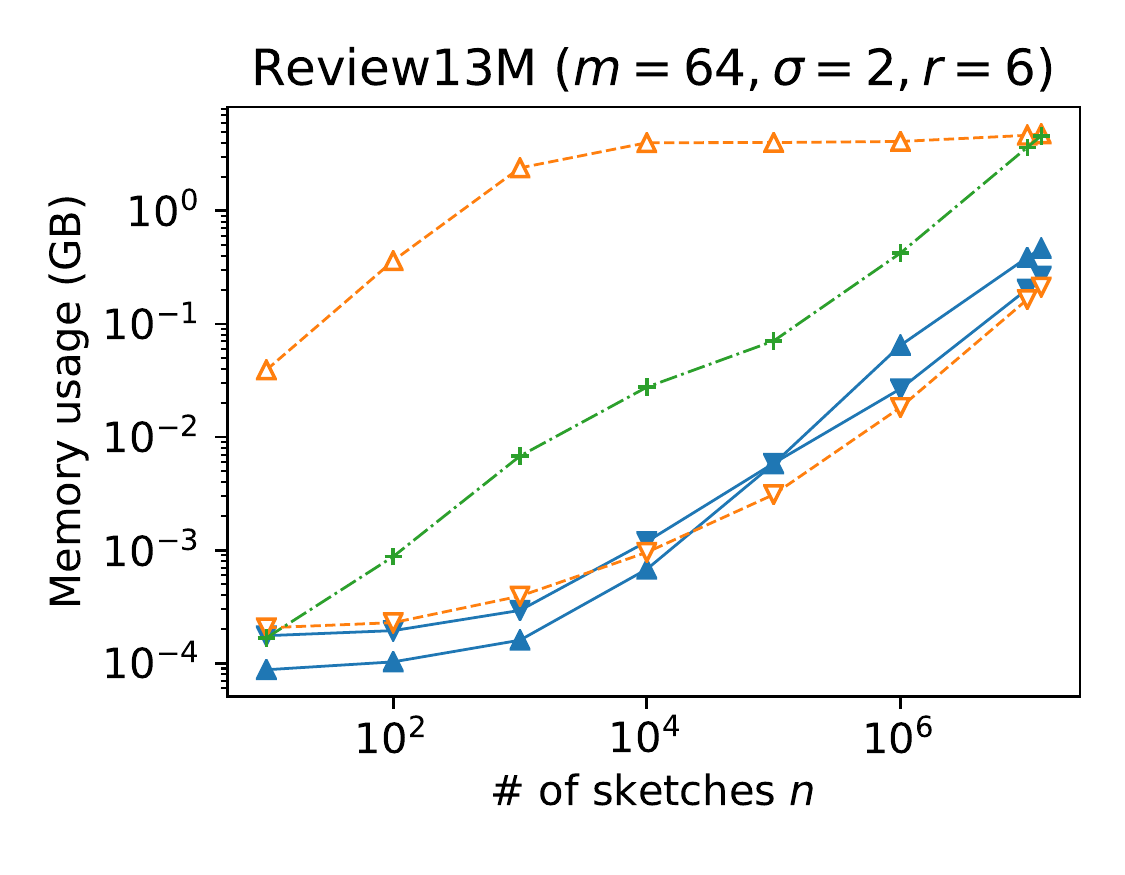}\\
\includegraphics[width=\ChartWidth]{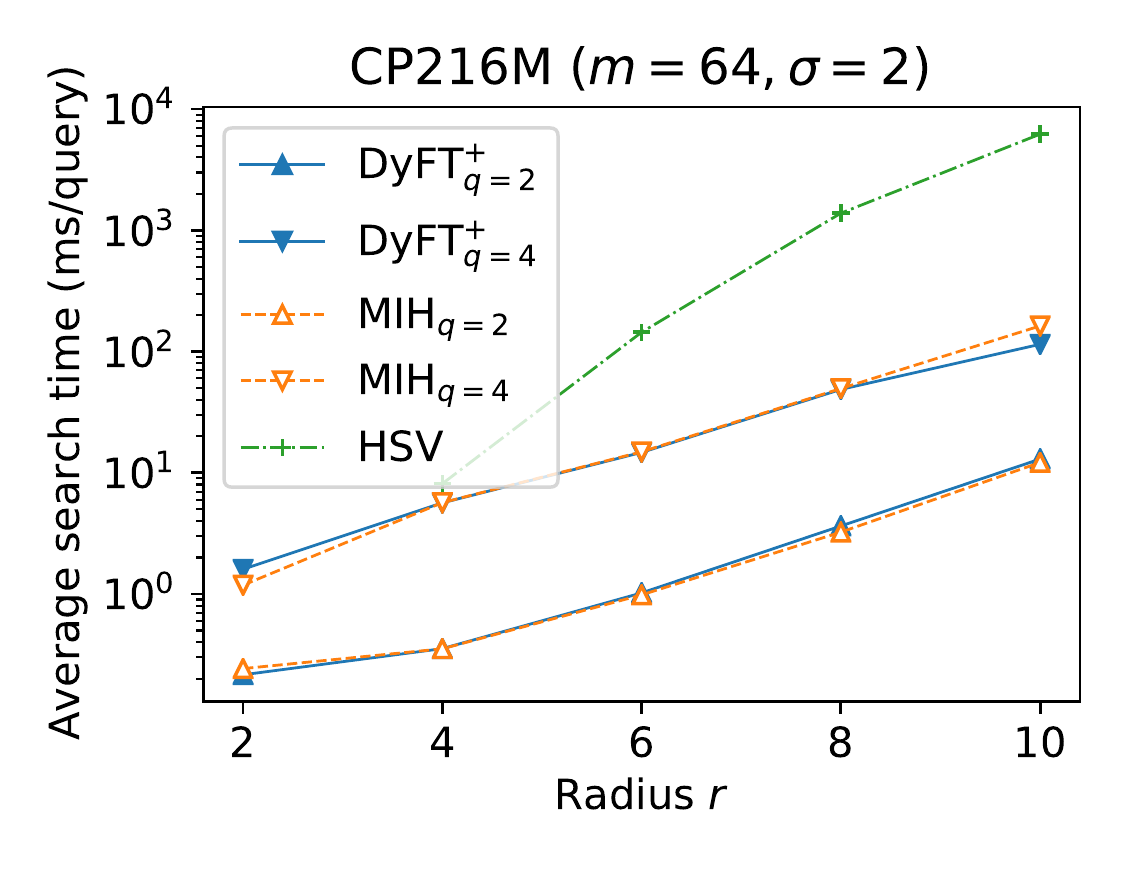}
\includegraphics[width=\ChartWidth]{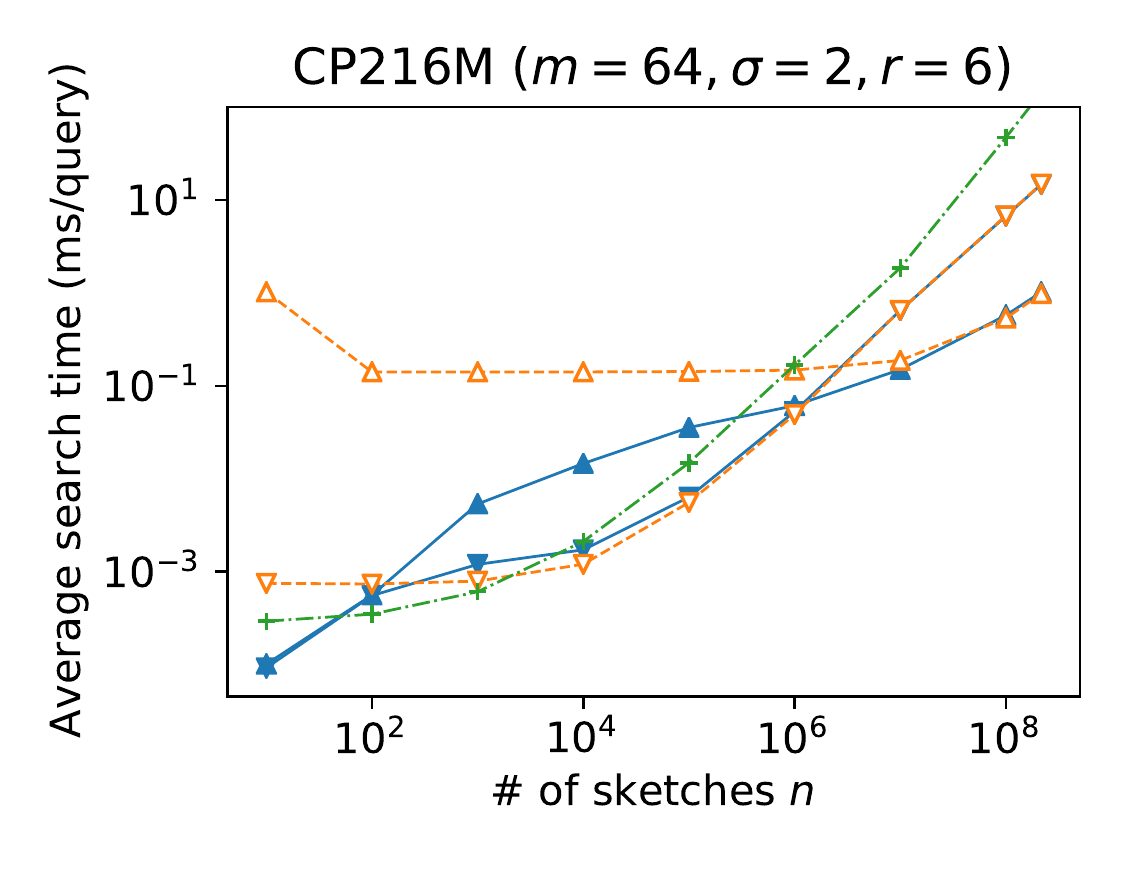}
\includegraphics[width=\ChartWidth]{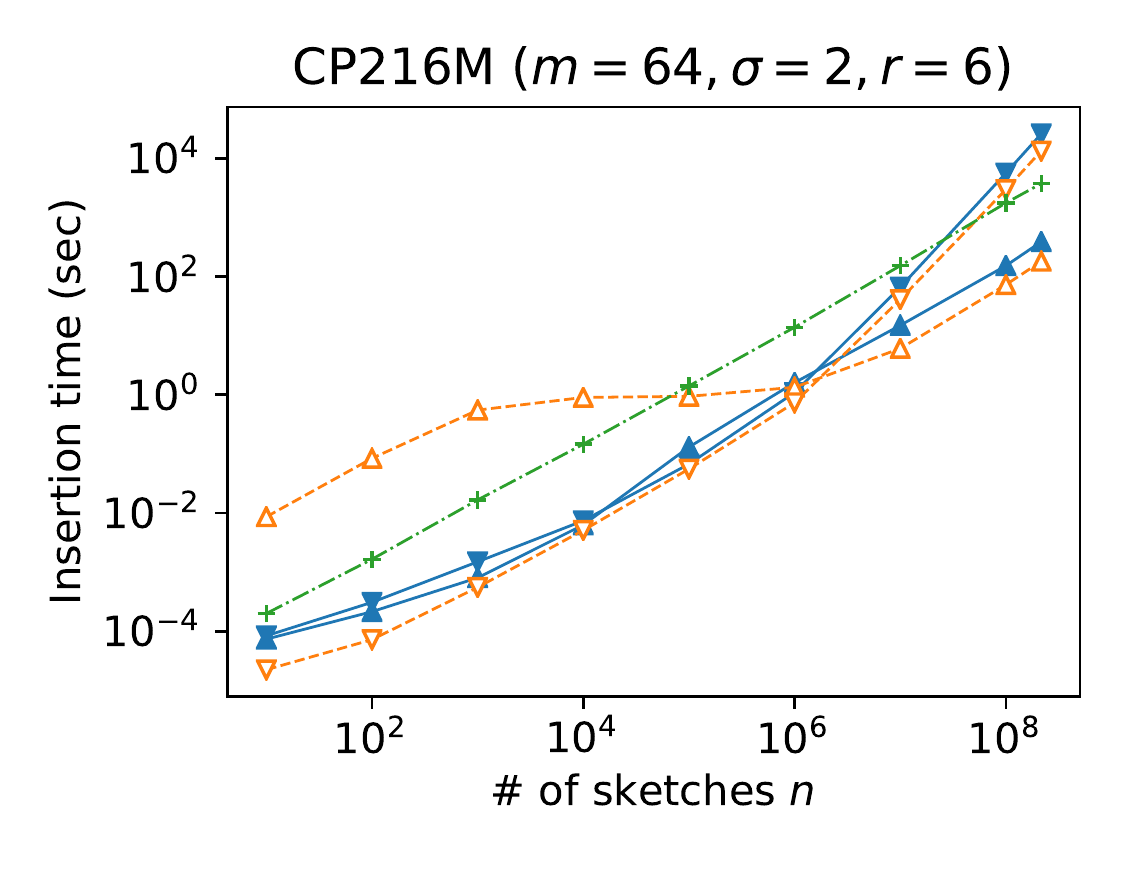}
\includegraphics[width=\ChartWidth]{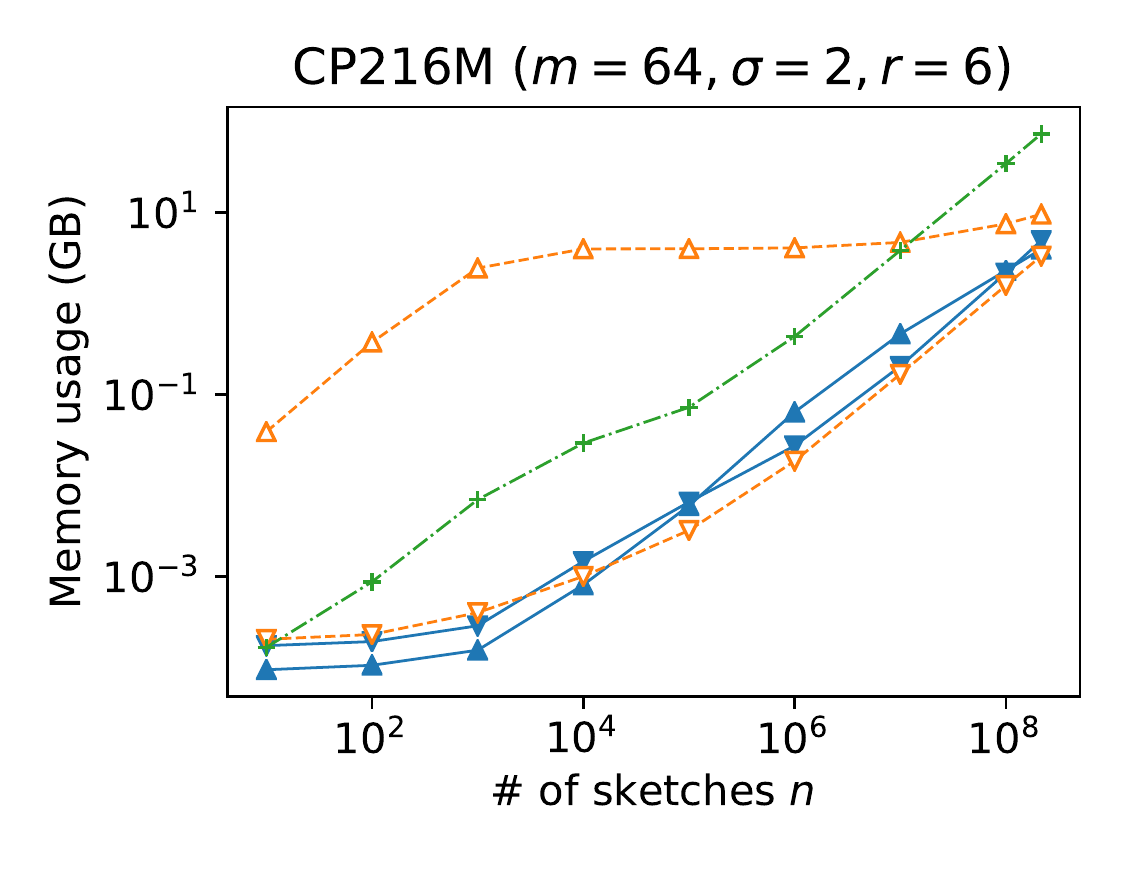}
\caption{
Comparison results for \ASFTP{}, MIH, and HSV on binary sketches.
The leftmost column shows average search time in milliseconds for varying radius $r$.
The second leftmost, third leftmost, and rightmost columns respectively show average search time in milliseconds, insertion time in seconds, and memory usage in GB, for varying the number of input sketches $n$.
The search time of HSV on \CP{} when $r = 2$ is not plotted since we could not construct the complete index within 256 GB of memory.
They are plotted in \textbf{logarithmic scale}.
}
\label{fig:asftp_binary}
\end{figure*}

We compared the performances of \ASFTP{}, MIH, and HSV on binary sketches.
MIH  is an early similarity search method using the multi-index approach \cite{norouzi2014fast}.
We implemented MIH using the original source code available at \url{https://github.com/norouzi/mih}.
HSV is a variant of HmSearch optimized for binary sketches \cite{zhang2013hmsearch}.
We implemented HSV applicable to dynamic settings using std::unordered\_map.
We tested $q = 2, 4$ for \ASFTP{} and MIH to observe the effect of the number of blocks on performance.
Note that the only difference between \ASFTP{} and MIH is whether a DyFT or hash-table structure is used to implement the index.

\fref{fig:asftp_binary} shows the results of search time, insertion time, and memory usage.
Since HSV was not competitive, we consider only on \ASFTP{} and MIH.
We first focus on the average search time for varying $r$ (on the leftmost column).
The search times of \ASFTP{} and MIH were not much different when all sketches in the dataset were inserted.
Both \ASFTP{} and MIH with $q = 2$ performed superiorly when the dataset had large $n$.
We now focus on the average search time for varying $n$ (on the second leftmost column).
As reviewed in \sref{sect:review}, the performance of MIH significantly degraded according to $n$.
MIH with $q=2$ was fast when $n$ was large, but very slow when $n$ was small.
\ASFTP{} maintained faster searches even when $n$ was small.
For insertion time and memory usage (on the two rightmost columns), MIH with $q=2$ was significantly worse when $n$ was small.
The result demonstrated that \ASFTP{} with $q=2$ is an excellent similarity search method if the dataset is dynamic and expected to be large.

\subsection{Analysis for \ASFTP{} on Integer Sketches}

\begin{figure*}[tb]
\centering
\includegraphics[width=\ChartWidth]{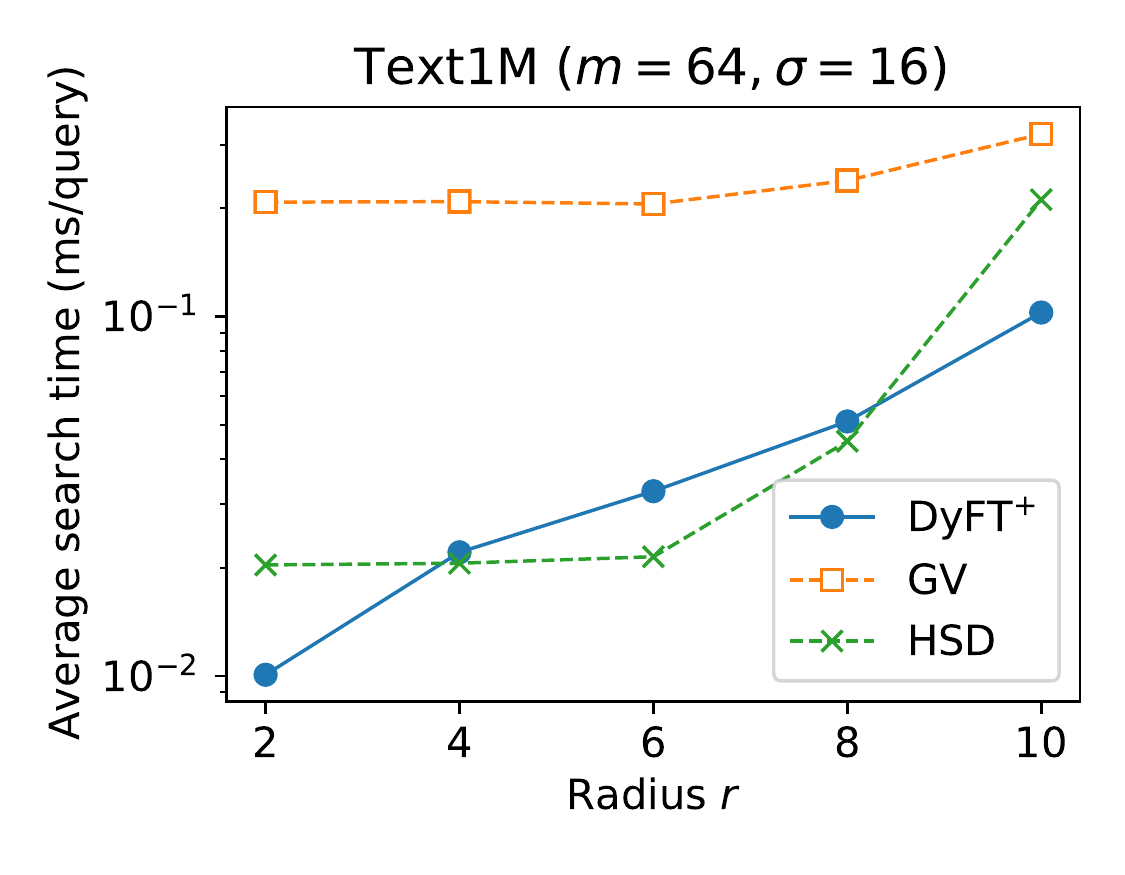}
\includegraphics[width=\ChartWidth]{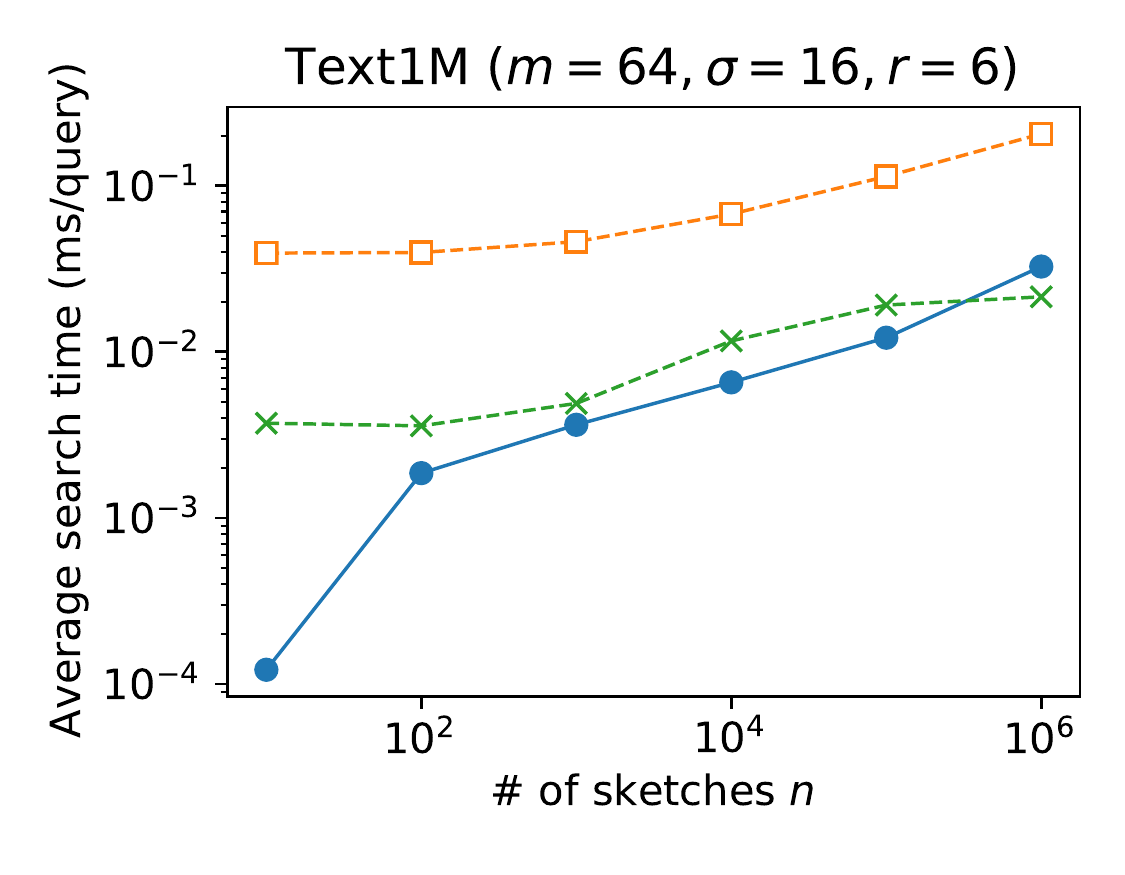}
\includegraphics[width=\ChartWidth]{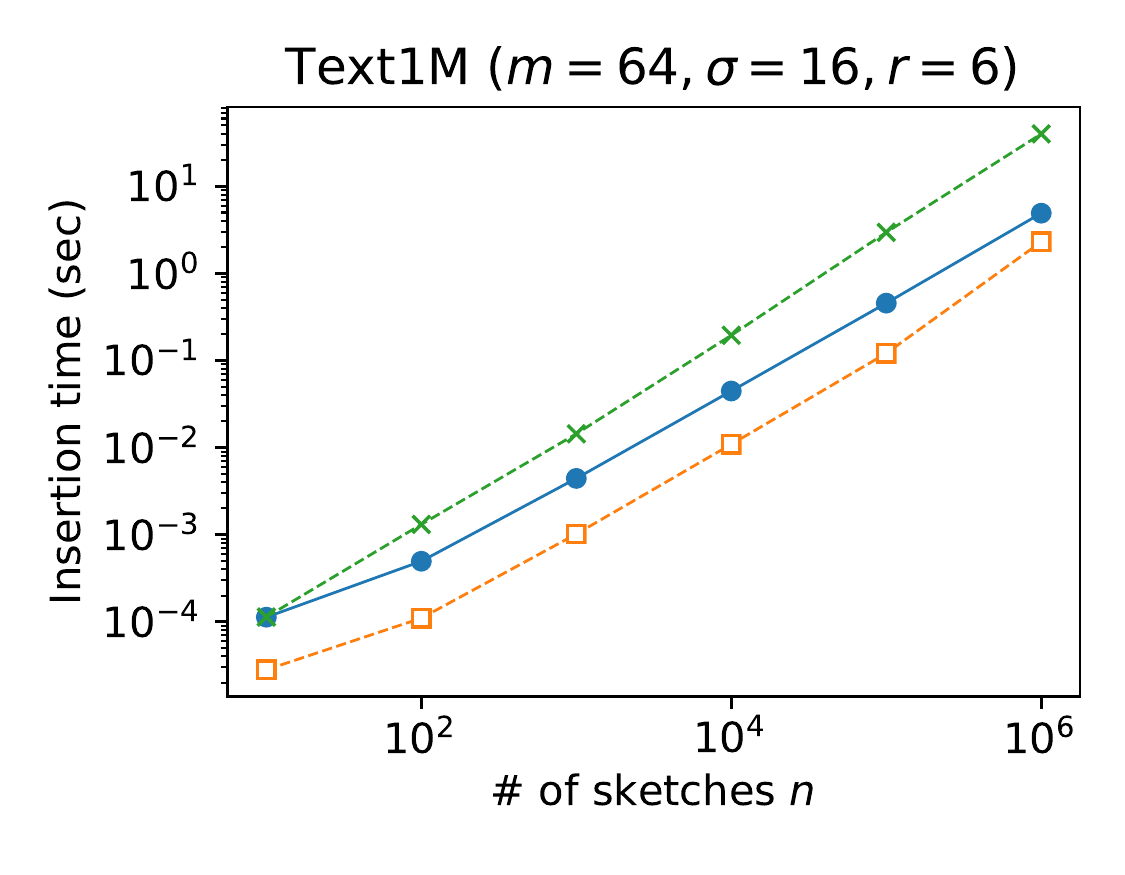}
\includegraphics[width=\ChartWidth]{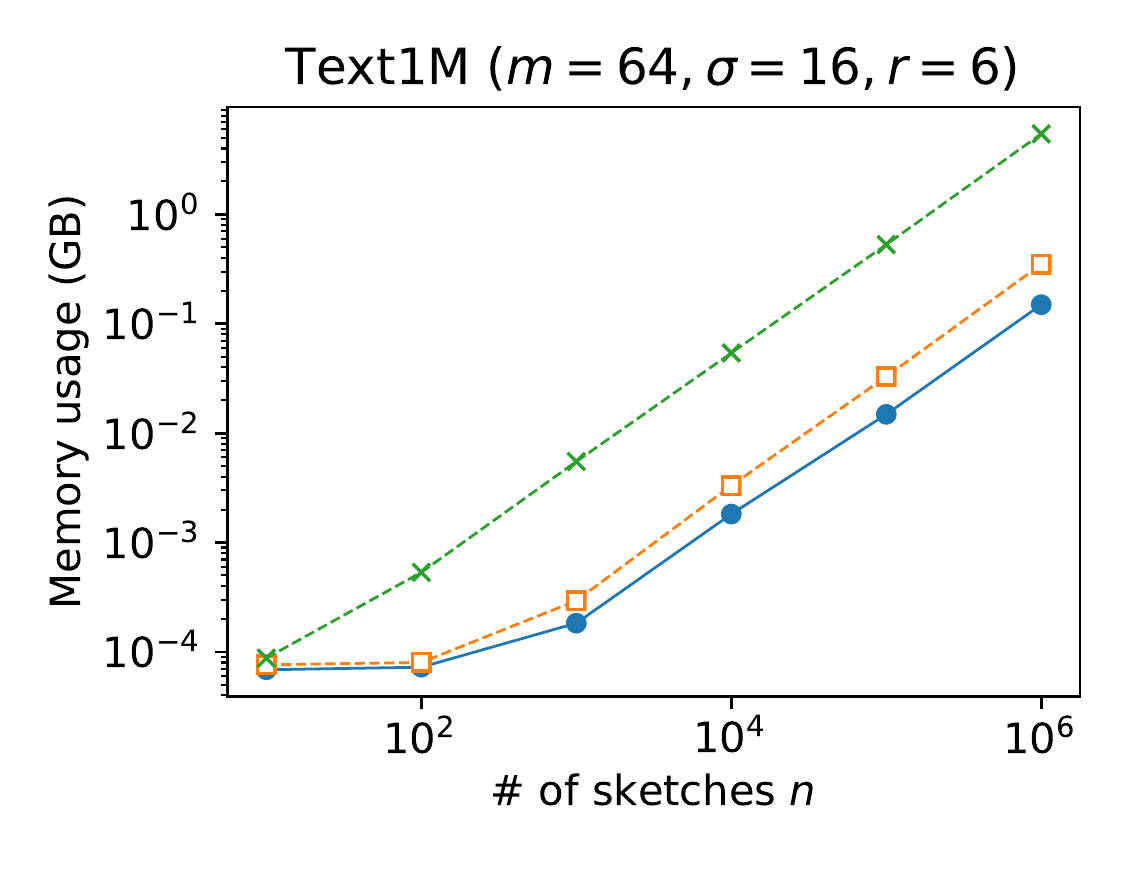}\\
\includegraphics[width=\ChartWidth]{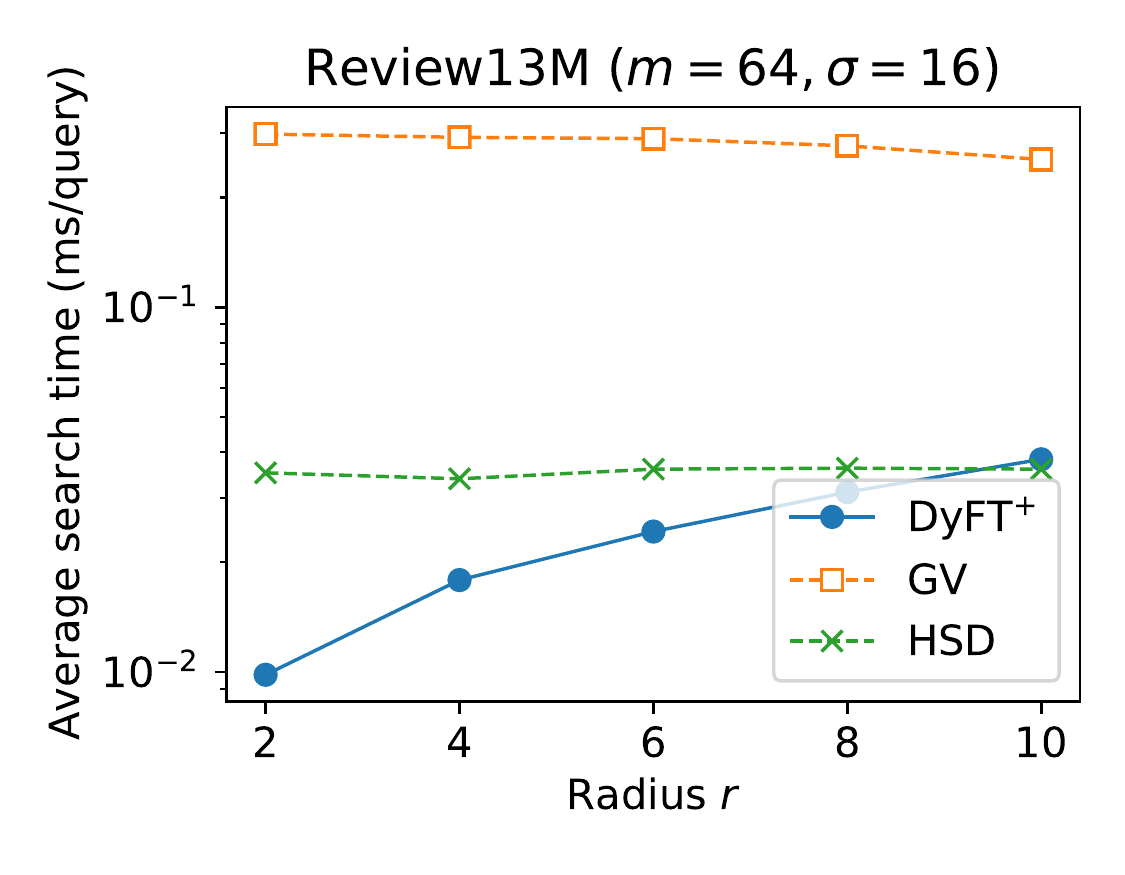}
\includegraphics[width=\ChartWidth]{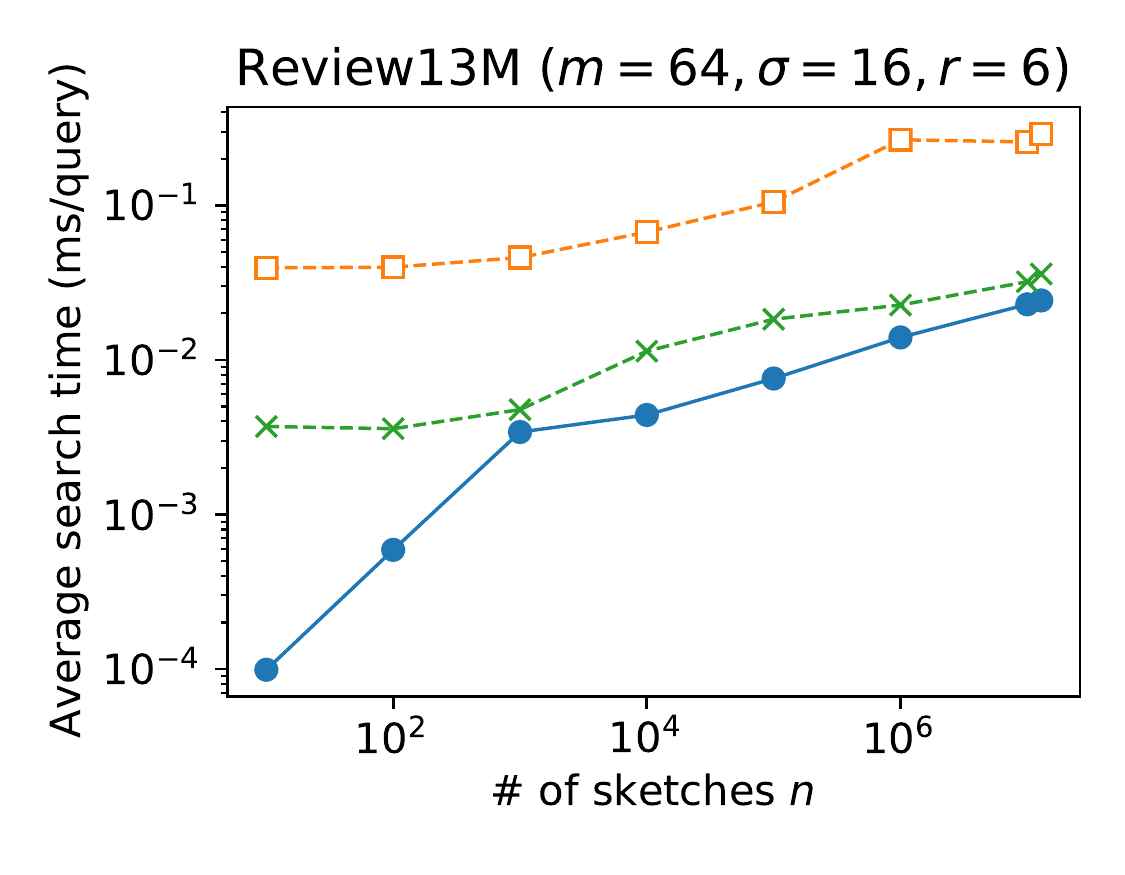}
\includegraphics[width=\ChartWidth]{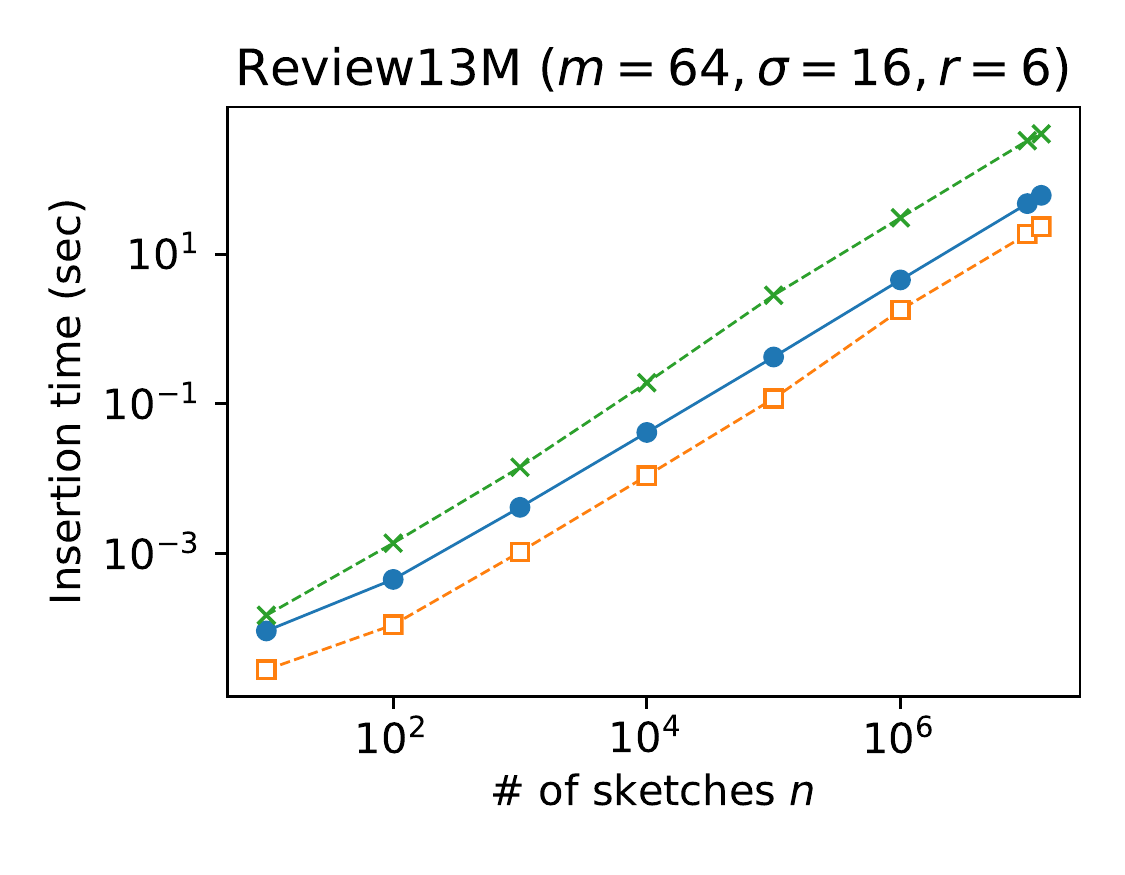}
\includegraphics[width=\ChartWidth]{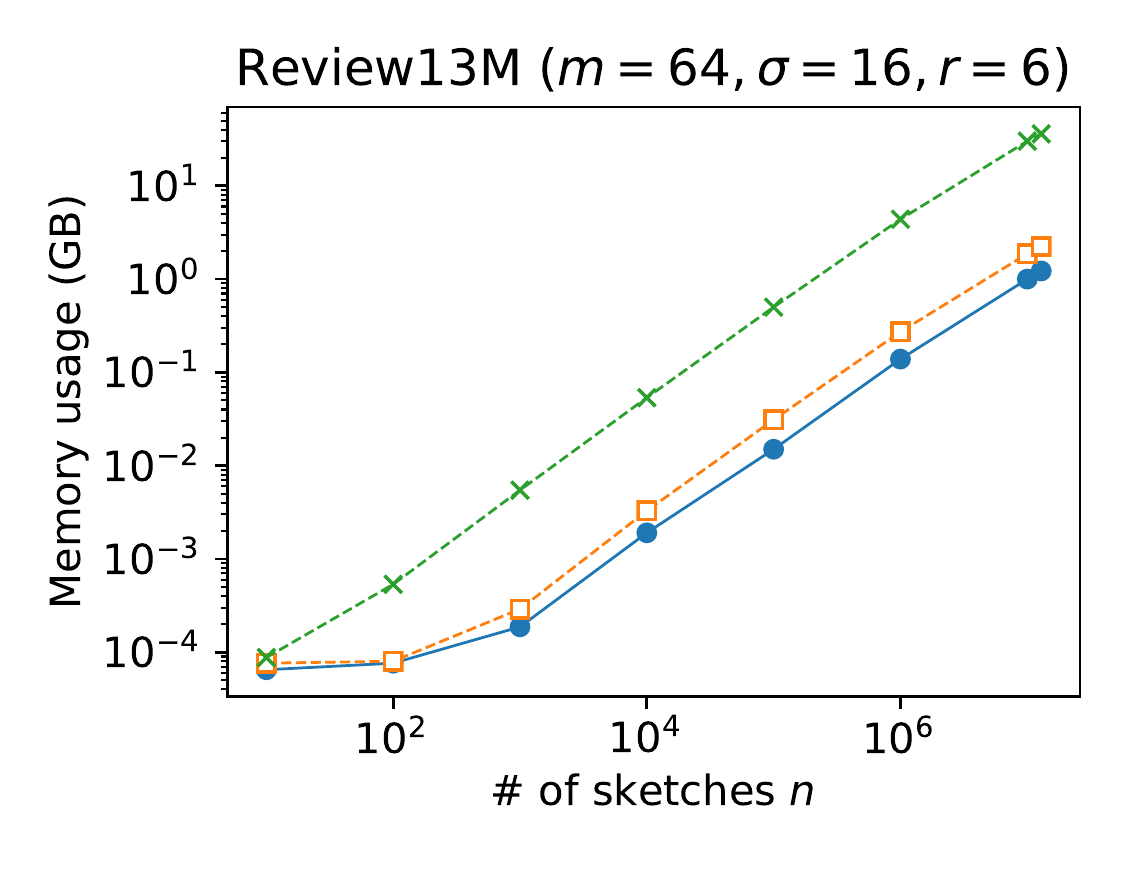}\\
\includegraphics[width=\ChartWidth]{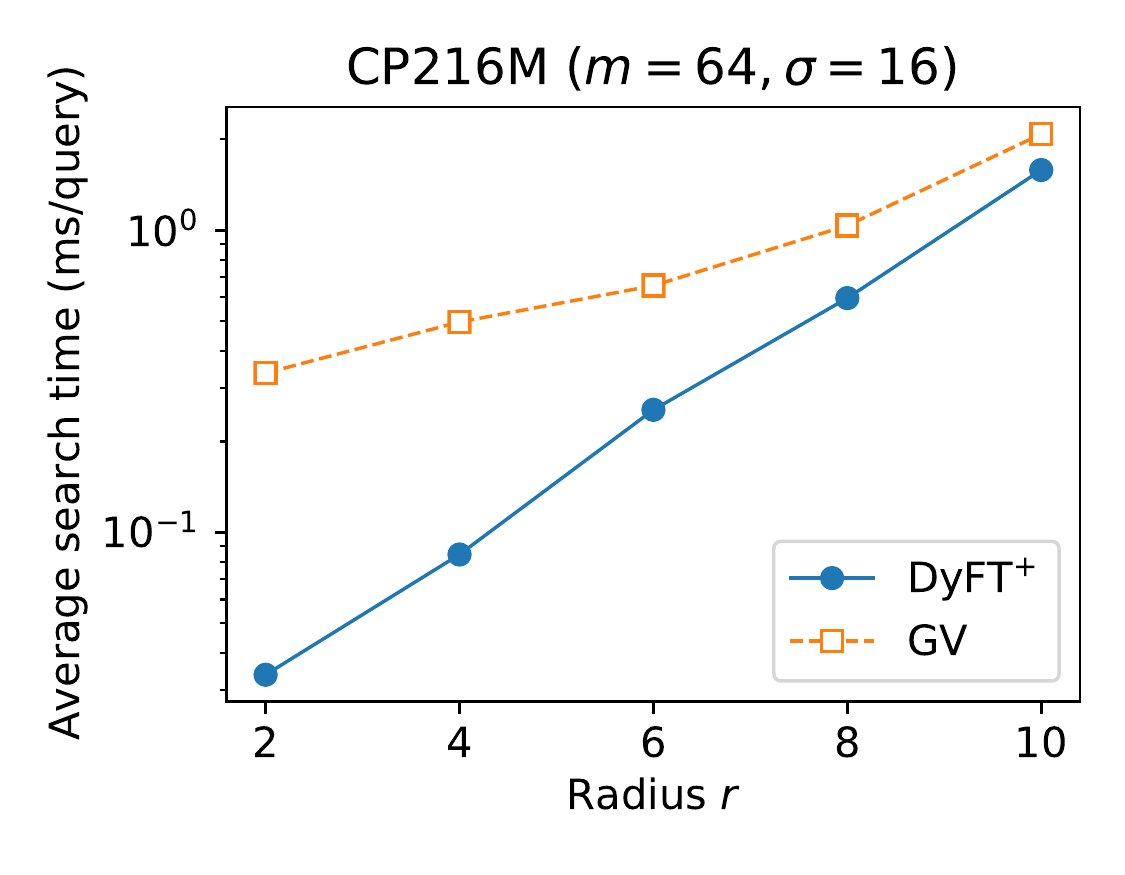}
\includegraphics[width=\ChartWidth]{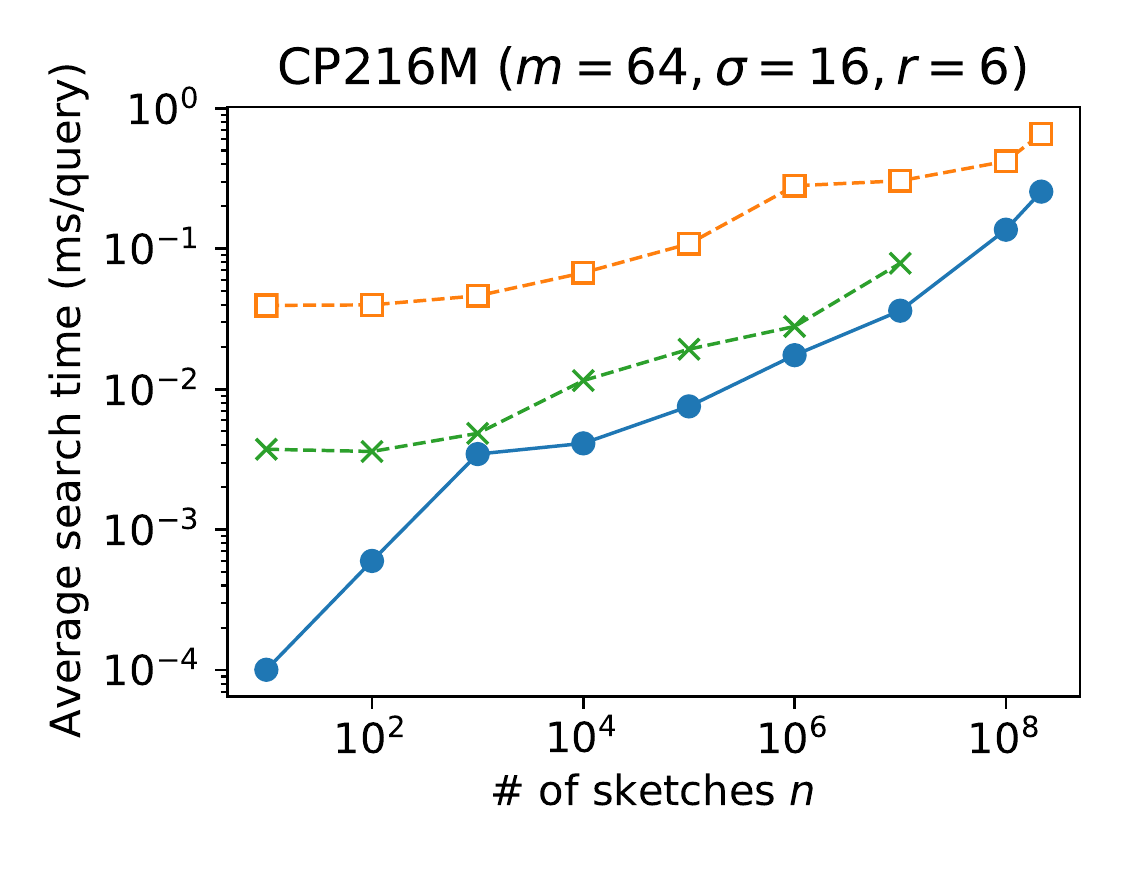}
\includegraphics[width=\ChartWidth]{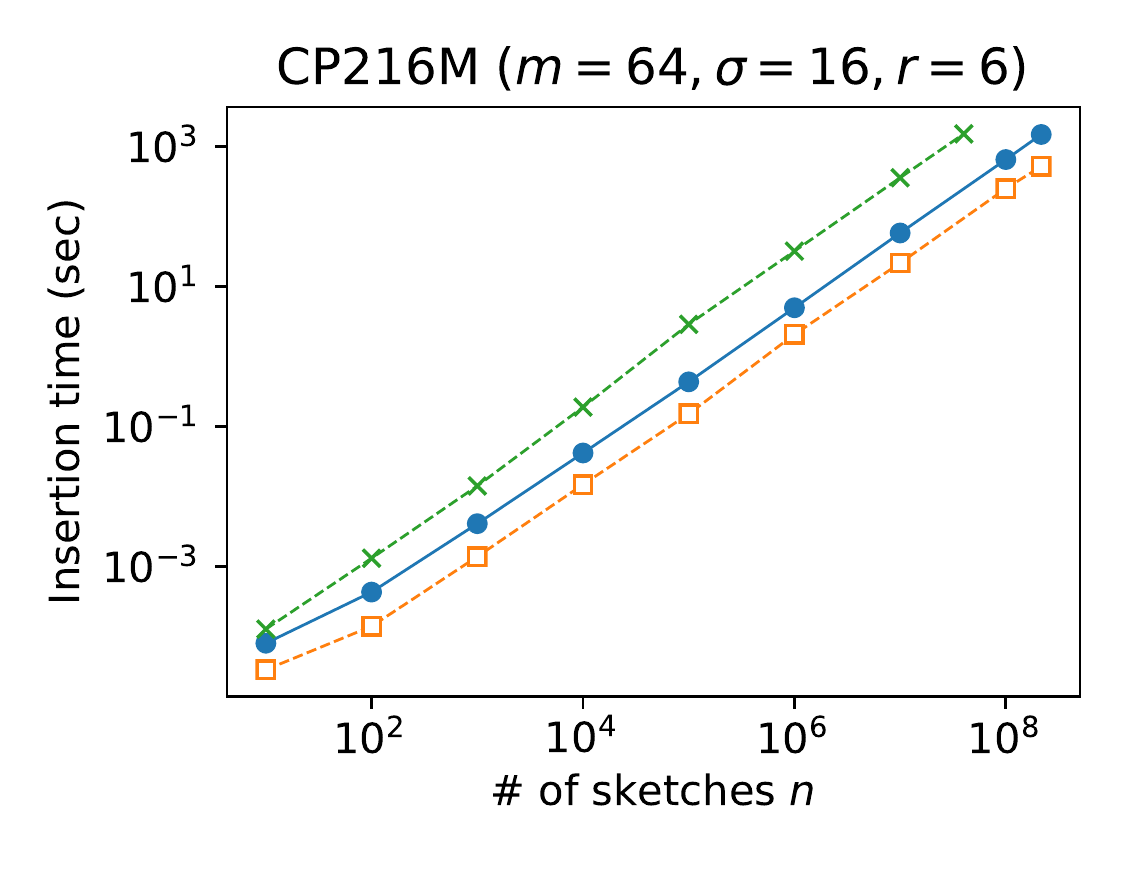}
\includegraphics[width=\ChartWidth]{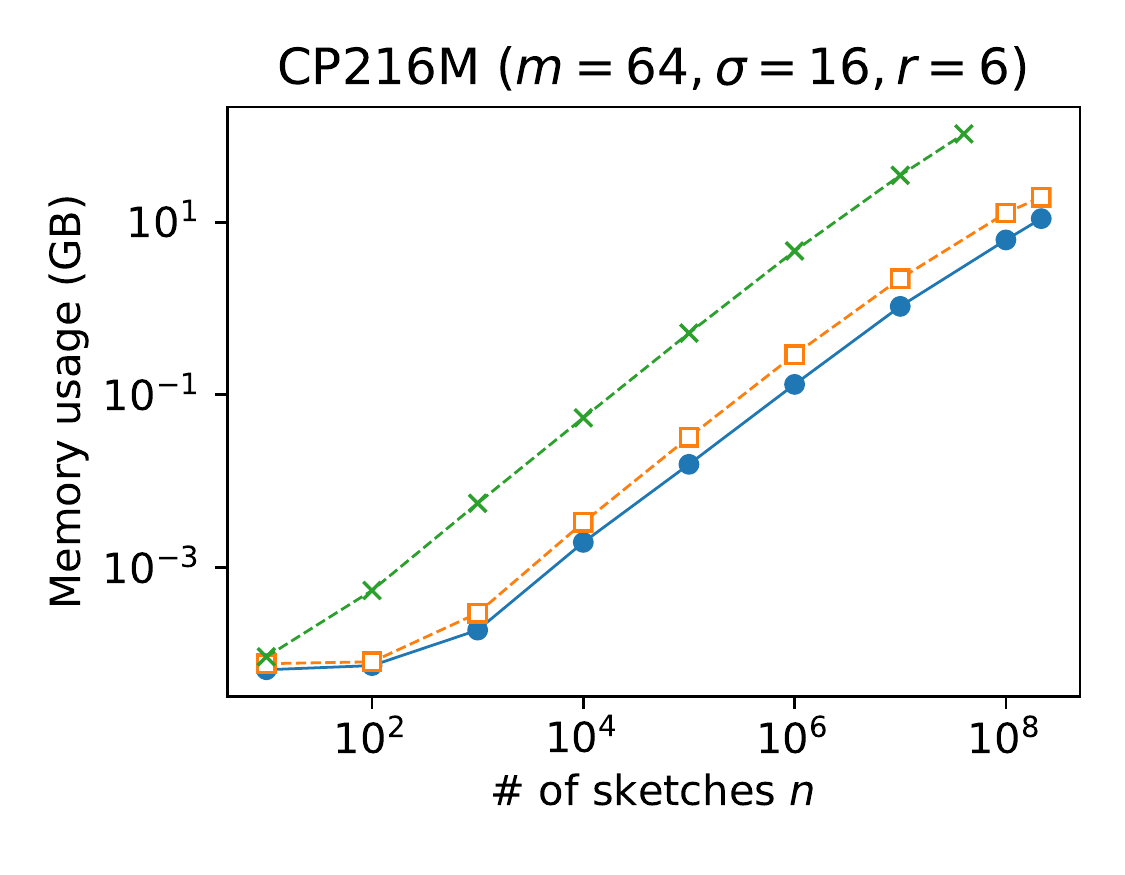}
\caption{
Comparison results for \ASFTP{}, GV, and HSD on integer sketches.
The leftmost column shows average search time in milliseconds for varying radius $r$.
The second leftmost, third leftmost, and rightmost columns respectively show average search time in milliseconds, insertion time in seconds, and memory usage in GB, for varying the number of input sketches $n$.
Some results of HSD on \CP{} are not plotted since we could not complete to construct the index within 256 GB of memory.
They are plotted in \textbf{logarithmic scale}.
}
\label{fig:asftp_integer}
\end{figure*}

We compared the performances of \ASFTP{}, GV, and HSD on integer sketches.
GV is a simple modification of MIH based on the idea in \cite{gog2016fast}. 
HSD is a variant of HmSearch optimized for integer sketches \cite{zhang2013hmsearch}.
We implemented GV and HSD applicable to dynamic settings using {std::unordered\_map}.
The only difference between \ASFTP{} and GV is whether a DyFT or hash-table structure is used to implement the index.
To fairly compare \ASFTP{} with GV, we set $q = \Floor{r/2}+1$ in \ASFTP{} in the same manner as GV.

\fref{fig:asftp_integer} shows the results of search time, insertion time, and memory usage.
We first focus on the average search time (on the two leftmost columns).
GV was not competitive to \ASFTP{} and HSD.
\ASFTP{} outperformed HSD in most cases.
We now focus on the insertion time and memory usage (on the two rightmost columns).
HSD was not competitive to \ASFTP{} and GV, as reviewed in \sref{sect:review}.
The insertion time of GV was the fastest because of its very simple data structure.
The memory usage of \ASFTP{} was the smallest because of the node-omitting approach and MART.
The result demonstrated that \ASFTP{} is a fast, scalable, and dynamic similarity search method on integer sketches.

\section{Conclusion}

We presented a dynamic similarity search method called DyFT and its multi-index variant called \ASFTP{} for the general Hamming distance problem.
Our experimental analyses using real-world datasets demonstrated that DyFT and \ASFTP{} outperform state-of-the-art similarity search methods.

\section*{Acknowledgments}
This work was supported by JST AIP-PRISM (grant number JPMJCR18Y5).
We thank the anonymous reviewers for their helpful comments.

\bibliographystyle{IEEEtran}
\bibliography{bibfiles/library_Lv3}

\end{document}